\documentclass[fleqn,usenatbib]{mnras}

\usepackage{newtxtext,newtxmath}

\usepackage[T1]{fontenc}

\DeclareRobustCommand{\VAN}[3]{#2}
\let\VANthebibliography\thebibliography
\def\thebibliography{\DeclareRobustCommand{\VAN}[3]{##3}\VANthebibliography}

\usepackage{graphicx}	
\usepackage{amsmath}	

\newcommand{\dmunits}{\ensuremath{\rm pc \, cm^{-3}}}

\newcommand{\dmhalo}{\ensuremath{\mathrm{DM}_\mathrm{halo}}}
\newcommand{\dmhalos}{\ensuremath{\mathrm{DM}_\mathrm{halos}}}

\newcommand{\dmfrb}{\ensuremath{\mathrm{DM}_\mathrm{FRB}}}
\newcommand{\dmigm}{\ensuremath{\mathrm{DM}_\mathrm{IGM}}}
\newcommand{\dmlu}{\ensuremath{\mathrm{DM}_\mathrm{LU}}}

\newcommand{\dmhost}{\ensuremath{\mathrm{DM}_\mathrm{host}}}

\newcommand{\dmmw}{\ensuremath{{\rm DM}_{\rm MW}}}
\newcommand{\dmmwhalo}{\ensuremath{{\rm DM}_{\rm MW,halo}}}
\newcommand{\dmmwism}{\ensuremath{{\rm DM}_{\rm MW,ISM}}}

\newcommand{\dmcosmic}{\ensuremath{\mathrm{DM}_\mathrm{cosmic}}}

\newcommand{\mhalo}{\ensuremath{M_\mathrm{halo}}}
\newcommand{\mstar}{\ensuremath{M_\mathrm{*}}}

\newcommand{\msun}{\ensuremath{{\rm M}_\odot}}

\newcommand{\figm}{\ensuremath{f_\mathrm{igm}}}

\newcommand{\fgas}{\ensuremath{f_\mathrm{gas}}}
\newcommand{\rmax}{\ensuremath{r_\mathrm{max}}}
\newcommand{\dm}{\ensuremath{\mathrm{DM}}}
\newcommand{\rvir}{\ensuremath{R_\mathrm{vir}}}

\newcommand{\Mpc}{\ensuremath{\rm{Mpc}}}

\newcommand{\hestia}{\textsc{Hestia}}
\newcommand{\hamlet}{\textsc{Hamlet}}
\newcommand{\ahf}{\textsc{Ahf}}

\usepackage[normalem]{ulem}

\title[Local Universe DM Model]{Modeling the Cosmic Dispersion Measure in the D < 120 Mpc Local Universe}

\author[Yuxin Huang et al.]
    {Yuxin Huang,$^{1}$\thanks{E-mail: mochafhxy@gmail.com}
    Khee-Gan Lee,$^{1,2}$ 
    Noam I. Libeskind,$^{3}$ 
    Sunil Simha,$^{4,5}$ 
    Aur\'{e}lien Valade,$^{6}$
    \newauthor
    J. Xavier Prochaska$^{1,7,8,9}$
\\
$^{1}$Kavli IPMU (WPI), UTIAS, The University of Tokyo, Kashiwa, Chiba 277-8583, Japan\\
$^{2}$Center for Data Driven Discovery, Kavli IPMU (WPI), UTIAS, The University of Tokyo, Kashiwa, Chiba 277-8583, Japan \\
$^{3}$Leibniz-Institut f\"ur Astrophysik Potsdam (AIP), An der Sternwarte 16, D-14482 Potsdam, Germany\\
$^{4}$Center for Interdisciplinary Research in Astrophysics, Northwestern University, 1800 Sherman Avenue, Evanston, IL 60201, USA\\
$^{5}$Department of Astronomy and Astrophysics, University of Chicago, William Eckhardt Research Center, 5640 S Ellis Ave, Chicago, IL 60637, USA\\
$^{6}$Aix Marseille Universit\'{e}, CNRS/IN2P3, CPPM, Marseille, France \\
$^{7}$Division of Science, National Astronomical Observatory of Japan, 2-21-1 Osawa, Mitaka, Tokyo 181-8588, Japan\\
$^{8}$University of California, Santa Cruz, 1156 High St., Santa Cruz, CA 95064, USA\\
$^{9}$Simons Pivot Fellow\\
}

\date{Accepted XXX. Received YYY; in original form ZZZ}

\pubyear{\the\year{}}

\begin{document}
\label{firstpage}
\pagerange{\pageref{firstpage}--\pageref{lastpage}}
\maketitle

\begin{abstract}
The Local Universe ($D< 120$\,Mpc) has been intensely studied for decades, with highly complete galaxy redshift surveys now publicly available.
These data have driven density reconstructions of the underlying matter density field, as well as constrained simulations that aim to reproduce the observed structures.
In this paper, we introduce a dispersion measure (DM) model that makes use of this detailed knowledge of our Local Universe within $D<120$ Mpc. The model comprises three key components: (i) the DM from the Milky Way's halo and the intra-group medium (up to 3.4 Mpc), derived from the HESTIA simulations, a series of constrained hydrodynamic simulations designed to reproduce our Local Group; (ii) the DM contribution from the large-scale intergalactic medium beyond the Local Group (3.4 Mpc $<D<120$ Mpc), calculated using the HAMLET reconstructed matter density field; and (iii) the individual DM contributions from Local Universe galaxy halos and clusters based on data from the 2MASS Galaxy Group Catalog and the NASA/IPAC Extragalactic Database. This comprehensive model will be made available as a Python package. As the most realistic model to date for DM in the local volume, it promises to improve the constraints of DM contributions from the Intergalactic Medium and Circumgalactic Medium of FRBs, thereby enhancing the accuracy of cosmic baryon distribution calculations based on DM analysis of FRBs.
\end{abstract}




\section{Introduction}\label{intro}

Fast Radio Bursts (FRBs) are one of the most intriguing astrophysical phenomena discovered in the last two decades. They are rapid and intense radio transients that last for milliseconds, and are thought to originate from extragalactic sources.
Since the first FRB detection in 2007 \citep{Lorimer2007}, various surveys and observatories have been instrumental in identifying and cataloging hundreds of FRBs. 
The Canadian Hydrogen Intensity Mapping Experiment (CHIME/FRB) \citep{CHIME2018,CHIME2021,Chime2023} focuses on wide-sky coverage and has detected a large number of FRBs, including many repeaters, thanks to its unique low-frequency range (400–800 MHz) and daily coverage on the entire northern sky.
On the other hand, surveys using interferometric radio arrays including the Commensal Real-time ASKAP Fast Transients (CRAFT) Survey \citep{Macquart2010,Shannon2024}, the Deep Synoptic Array (DSA) \citep{Ravi2023DSA,Law2024} and MeerTRAP \citep{Jankowski2022,Rajwade2022} emphasize real-time detection and localization
of smaller numbers of non-repeaters to approximately one
arcsecond precision.

The dispersion measure (DM) of FRBs is a key observable that quantifies the total column density of free electrons along the line of sight between the FRB source and the observer. It is defined as:
\begin{equation}
    \dm \equiv \int \frac{n_e}{1 + z} \, \mathrm{d}l
\end{equation}
where $n_e$ is the free electron number density and $\mathrm{d}l$ is the proper distance element. 
The total DM is usually divided into 3 components:
\begin{equation} \label{eq:dm_components}
    \dm=\dmmw+\dmcosmic+\frac{\dmhost}{1+z}
\end{equation}
where \dmmw{} is the DM contributed by the Milky Way (MW), $\dmcosmic=\dmigm+\dmhalos$ is the diffuse cosmic contribution, which we will further decompose into the intergalactic medium (IGM), i.e. from gas within filaments and voids, and the foreground halo contributions. Finally, \dmhost{} is the DM contributed by the host galaxy and the local medium in the immediate vicinity of the FRB source. 

FRBs have been proven to be powerful tools for addressing some of the most profound questions in cosmology, utilizing the \dmcosmic{} component which carries information of ionized gas inside the IGM and circumgalactic medium (CGM). One of the key contributions of FRBs is their role in resolving the long-standing problem of the ``missing baryons'' in the universe \citep{Fukugita2004,Cen2006,Bregman2007}. The Macquart relation \citep{Macquart2020}, which demonstrated that the $\dm(z)$ of localised extragalactic FRBs can account for the expected cosmic baryon content along their line of sight, has heralded FRBs as a useful probe of the cosmic baryonic gas, particularly those in the warm-hot intergalactic medium (WHIM). 
\citet{Simha2020} demonstrated the next logical step by using foreground spectroscopic data to demonstrate that the various components in Equation~\ref{eq:dm_components} was in general agreement with the value measured from the FRB.
The ongoing FLIMFLAM survey \citep{Lee2022,Khrykin2024a,Huang2024} aims to collect similar foreground data sets on a sample of localised FRBs to place constraints not only on the cosmic baryon fraction within the IGM but also on the ionized gas fraction inside halos.

Moreover, FRBs enable new and independent estimates of the Hubble constant $H_0$ \citep{Wu2022,Hagstotz2022,James2022,Gao2024}, complementing traditional methods like the cosmic microwave background (CMB) analysis and Type Ia supernovae (SNe) observations. It is predicted that $H_0$ can be constrained to within $\approx4\%$ using 100 well-localized FRBs \citep[][but see \citet{Baptista2024}]{James2022}, which could be reached in the following the CRAFT COherent upgrade (CRACO) observations and CHIME outrigger project \citep{Leung2023}. 
FRBs can also constrain the cosmological parameters in conjunction with CMB, baryon acoustic oscillation (BAO), SNe and gravitational waves \citep{Walters2018,Wei2018}. \citet{Shirasaki2022} predicted that the cross-correlation between the dark matter halos and the \dmcosmic{} of 20,000 FRBs with a localization error being 3 arcmin are able to constrain the cosmological parameters to 5\%. In addition to their cosmological applications, FRBs provide a unique probe of structure formation and galactic feedback by analyzing deviations of \dmcosmic{} from the mean \dmcosmic{} \citep{Batten2022,Baptista2024}. The standard deviation of this distribution correlates with feedback strength and the redshift, allowing for constraints on different feedback models in hydrodynamic simulations \citep{Medlock2024}. 
Since both the gas partition between the IGM and halos, and the hot gas fraction within each halo are sensitive to feedback mechanisms, comparing the constraints from the observations with hydrodynamic simulations across different feedback models can further refine our understanding of these mechanisms \citep{Khrykin2024b}.

However, the aforementioned studies only utilize \dmcosmic, with \dmmw{}
considered as a nuisance parameters. So it is essential to model the DM contributed by the MW precisely. 
Previously, \citet{Dolag2015} predicted the MW halo's DM contribution to be approximately 30 \dmunits{}, using a representative MW-type galaxy in the hydrodynamic cosmological simulation introduced by \citet{Beck2013}.
Nowadays the most widely used estimation of the MW halo contribution is $\dmmwhalo\approx50$ \dmunits with an uncertainty of 15 \dmunits{} as estimated by \citet{Prochaska2019}. In \citet{Prochaska2019}, the \dmmwhalo{} is separated into the cool halo component ($T\sim10^4$K) and the hot halo component ($T\gtrsim10^6$K). The DM of the cool Galactic halo is modeled with the high-velocity clouds detected by the all-sky 21 cm emission map observed by the HI4PI survey \citep{HI4PI2016}, while the DM of the hot Galactic halo is measured from the column density map of the highly ionized oxygen ions observed by the X-ray \citep{Fang2015} and far-UV surveys \citep{Sembach2003}.
Most recent works have generated other models
\citep{Yamasaki2020,Das+2021} or used FRB measurements
to constrain \dmmwhalo\ 
\citep{Platts2020,Cook2023,Ravi2023}.
\cite{Cook2023} estimates the upper limit of the \dmmwhalo{} as a function of the Galactic latitude utilizing low-DM CHIME FRBs. \cite{Ravi2023} investigated FRB20220319D, whose host galaxy is located at 50 Mpc. The presence of two nearby pulsars in the sky constrains the \dmmwhalo{} at this position, potentially
ruling out several MW halo DM models \citep{Maller2004,Faerman2017,Prochaska2019}. The DM of those 2 pulsars provide an upper limit of the \dmmwhalo{} at the FRB coordinate. While all these studies produce empirical constraints, they are limited by the on-sky density of sightlines that sample the halo. 

To date, the community has largely ignored
an additional source of variance to analyses of 
\dmcosmic: the DM contributed from 
structures in the Local Universe.
\cite{Prochaska2019} noted that M31 and 
the Milky Way satellites (e.g.\ the Magellanic
clouds) may impose DM of several tens
to over one hundred \dmunits.
In addition, the Local Group may harbor
an intragroup medium (IGrM) with a highly
asymmetric signal.  In this paper, we 
also consider one additional source:
the cosmic web of the Local Universe.

Over multiple decades, our understanding of the Local Universe has been significantly advanced by numerous imaging and spectroscopic surveys, which offers insights into the distribution of matter and the dynamics of galaxies within a few hundred megaparsecs. Spectroscopic surveys like the Sloan Digital Sky Survey (SDSS, \citet{Abazajian2009}, 6dF Galaxy Survey \citep{Jones2009} and the Two Micron All Sky Survey (2MASS, \citet{Jarrett2000}) have been crucial in mapping the Local Universe, revealing the large-scale distribution of galaxies and cosmic structures. The Tully 2MASS group catalog \citep{Tully2015}, derived from 2MASS, has identified key galaxy groups and clusters, enhancing our understanding of local cosmic structure and dark matter distribution. The NASA/IPAC Extragalactic Database Local Volume Sample \citep[NEDLVS][]{CookDO2023} has further enriched this understanding by synthesizing the aforementioned datasets along with redshift-independent distances for galaxies within $\sim40$ Mpc. On the other hand, peculiar velocity catalogs, like the Cosmicflows compilations \citep[for the latest release, CF4]{Tully2023}, can directly probe the spatial distribution of gravitating matter (luminous and dark), helping in identifying the influence of massive structures like the Virgo Cluster \citep{Blakeslee2009}, the Great Attractor \citep{Dressler1987}, and the Shapley Supercluster \citep{Raychaudhury1989} on galaxy motions and allows for better modeling of the dark matter distribution in the universe. The synergy between these imaging and spectroscopic surveys has been pivotal in enhancing our understanding of the Local Universe, offering a detailed view of its structure, composition, and dynamics. This wealth of data also serves as a foundation for more complex models and simulations, which further our understanding of the universe on cosmological scales. 

Despite the wealth of information provided by spectroscopic surveys and constrained simulations of the $D\lesssim 120$ Mpc Local Volume that cover the full sky, a comprehensive DM model that takes into account these data is still lacking. In this study, we devise a DM model of not only the MW halo and Local Group, but the whole Local Universe within a distance $D<120$ Mpc. We will construct a series of HEALPix projection maps for four distinct components: the MW halo, the Local Group intra-group medium (IGrM), the Local Universe IGM, and the Local Universe galaxy halos. For the the latter two components, we will provide projection models across 13 distance layers, each with a thickness of 8 Mpc. 
We have made this model publicly available as apython package \texttt{pyhesdm}\footnote{\href{https://github.com/yuxinhuang1229/pyhesdm}{https://github.com/yuxinhuang1229/pyhesdm}}, which can compute the DM contribution from the different components at a given Galactic latitude and longitude.

This paper is organized as follows: In Section \ref{sec:sim} we briefly introduce the \hestia{} constrained hydrodynamical simulation, which we utilized to create our Local Group DM model. Section \ref{sec:mw} describes our DM model for the Milky Way, while Section \ref{subsec:igrm} covers the DM model of the Intra-group Medium (up to 3.4 Mpc). In Section \ref{sec:localU} we present the DM model components for the IGM and halos beyond the Local Group (3.4 Mpc$<D<120$ Mpc). Finally, in Section \ref{sec:discussion} we compare our model with previous studies to test its consistency with observations.

We use the cosmological parameters from \citet*{Planck2014}, ensuring consistency with the \hestia{} simulation: $\sigma_8 = 0.83$, $h = 0.677$,  $\Omega_\Lambda = 0.682$, $\Omega_{\rm{m}}=0.27$ and $\Omega_{\rm{b}}=0.048$.

\section{Simulations}\label{sec:sim}

\begin{table*}
	\centering
	\caption{The halo mass and virial radius of the simulated MW and M31 in the 37\_11 simulation are presented. For comparison, observational estimates of \mhalo{} are also provided. The estimated \mhalo{} for the MW listed here is based on proper motion measurements of globular clusters from the Gaia Data Release 2 catalog and observations from the Hubble Space Telescope, and the estimate for M31 \mhalo{} listed here is derived from its photometric data obtained through SDSS observations.}
	\label{tab:mw_m31_info}
	\begin{tabular}{lcccc} 
		\hline
		  & \mhalo & \rvir & Observed \mhalo & Reference \\
		\hline
		MW  & $1.02\times10^{12}\msun$ & 212.10 kpc & $1.3\pm0.3\times10^{12}\msun$ & \citet{Posti2019} \\
		M31 & $1.04\times10^{12}\msun$ & 213.60 kpc & $0.95\pm0.15\times10^{12}\msun$ & \citet{Tamm2012} \\
		\hline
	\end{tabular}
\end{table*}

The major part of our model, including the MW halo and the IGrM, uses the High-resolution Environmental Simulations of The Immediate Area (\hestia{}) simulation suite \citep{Libeskind2020}. This simulation is a set of cosmological hydrodynamic simulations designed to simulate the local cosmography. 

The initial conditions for the simulation are derived using real observational data, particularly from the CosmicFlows-2 catalog \citep{Tully2013}. This catalog includes over 8,000 direct distance measurements, which provide information about the peculiar velocities of galaxies within the Local Universe. Wiener Filter (WF) and Constrained Realizations (CR) methods \citep{Hoffman1991,Zaroubi1995} are then employed to reconstruct 1000 realizations of the initial densities and velocity fields of the Local Universe. These techniques ensure that the simulation's end point closely matches the actual observed universe. Before advancing to full hydrodynamic simulations, the \hestia{} suite begins with a dark-matter-only run, which is conducted in $100\text{ Mpc}h^{-1}$ periodic boxes with $256^3$ DM particles. 
Halos are then identified using a two-step process. The Friends-of-Friends (FoF) algorithm is first applied on the particle field, followed by the halo and subhalo identification by the \ahf{} halo finder\footnote{\href{http://popia.ft.uam.es/AHF/Download.html}{http://popia.ft.uam.es/AHF/Download.html}} \citep{Knollmann2009}.
Realizations featuring halo pairs analogous to the Milky Way and Andromeda galaxies, positioned near (SGX, SGY, SGZ) $= (0, 0, 0)$, with halo masses between [$8\times10^{11}\msun$, $3\times10^{12}\msun$] and separations within [0.5, 1.2] Mpc, alongside a massive halo representative of the Virgo Cluster ($\mhalo>2\times10^{14}\msun$) located approximately at (SGX, SGY, SGZ) $= (-3.5, 16.0, -0.8)$ Mpc, are selected for high-resolution hydrodynamic simulations.

Once the appropriate halo pairs are identified, a smaller region with approximately 3.5 Mpc around the Local Group is then re-simulated at $16^3$ or $32^3$ times greater resolution incorporating hydrodynamics and galactic feedback, using the Arepo code\footnote{\href{https://arepo-code.org}{https://arepo-code.org}} \citep{Weinberger2020}. These simulations include dark and baryonic matter while incorporating the Auriga galaxy formation model \citep{Grand2017}, which is broadly similar to IllustrisTNG \citep{Pillepich2018}. The model includes detailed prescriptions for physical processes such as gas cooling, star formation, feedback from supernovae, and the effects of magnetic fields. These simulations take into account the local environment's influence, such as the gravitational effects of nearby large structures like the Virgo Cluster and the dynamics of the Local Void.

The hydrodynamic simulations are conducted at two different resolutions. The intermediate resolution runs are at $4096^3$ effective particles within a $5\text{ Mpc}h^{-1}$ sphere. The mass and spatial resolution is $m_{\rm{dm}} = 1.2\times10^6\msun$, $m_{\rm{gas}} = 1.8\times10^5\msun$ and $\epsilon = 340\text{ pc}$.
The high-resolution runs achieve a resolution of $8192^3$ effective particles within a $2.5\text{ Mpc}h^{-1}$ sphere around the center of the Local Group. The mass and spatial resolution is $m_{\rm{dm}} = 1.5\times10^5\msun$, $m_{\rm{gas}} = 2.2\times10^4\msun$ and $\epsilon = 220\text{ pc}$. Among the three high-resolution runs, labeled 09\_18, 37\_11, and 17\_11, we selected the 37\_11 run to generate our model. This choice was made based on the disk scale of the simulated MW, determined from its surface brightness profile. The simulated MW in the 37\_11 run has a disk scale length that most closely matches observational data among the three simulations \citep{Bland-Hawthorn2016}. This match allows for a smooth transition between the NE2001 model and the \hestia{} simulation, which we used to model the DM of the MW disk and halo, respectively (see Section \ref{sec:mw} for details). The halo mass and virial radius of MW and M31 in this simulation are presented in Table \ref{tab:mw_m31_info}.

\section{The Milky Way}\label{sec:mw}

We separate the DM contribution of the Milky Way into the disk and halo components, with the latter considered to occupy the region outside the Milky Way disk but within the sphere corresponding to one virial radius of the MW. 
For the Milky Way contributions, we only provide projected models, computing the DM from the Sun to the edge of the model.

\subsection{Interstellar Medium}\label{subsec:mwism}
Since we do not expect \hestia{} to have the resolution and accuracy to model the Milky Way disk contribution, we will instead adopt the empirical NE2001 \citep{Cordes2003} model for the MW's interstellar medium (ISM).
This is a widely used empirical model designed to predict the distribution of free electrons in the Galactic disk mainly based on the DM measured from Galactic pulsars. The model divides the MW into several components, estimating the ISM, thin disk, thick disk, spiral arms, Galactic center, clumps and voids components separately.
We use the NE2001 model instead of the more recent YMW16 model \citep{YMW16} because YMW16 has been shown to overestimate the Galactic DM toward the Galactic anti-center \citep{Price2021}. However, users of \texttt{pyhesdm} can easily switch to other models for the MW disk, as the DM contributions from different components are provided separately by the algorithm.

In our overall model, the MW ISM component is evaluated within a cylindrical disk with height 1 kpc and radius 20 kpc. 
The map is calculated with \texttt{pygedm}\footnote{\href{https://github.com/FRBs/pygedm}{https://github.com/FRBs/pygedm}} \citep{Price2021}, which is a Python package for calculating galactic electron density based on models such as NE2001 and YMW16. 
We do not make any modifications to the DM contributions derived from NE2001, but incorporate it as part of our software package for convenience.

\subsection{Galactic Halo}\label{subsec:mwhalo}

The DM model of the MW halo is derived from one of the three highest-resolution \hestia{} hydrodynamic simulations we have available, which is labeled 37\_11. We begin by outlining our method for calculating DM within the \hestia{} simulations. 
Given the coordinates of the start and end points of the sightline along which the DM is calculated, we divide the sightline into sufficiently small segments, ensuring that the length of each segment matches the scale of the smallest gas cell in the simulation. 
We then identify the closest gas cell to each line segment and assign its electron number density $n_{\rm{e}}$ to that segment. The proper electron number density is calculated by 
\begin{equation}
    n_{\rm{e,prop}}=e_{\rm{ab}}X_{\rm{H}}\frac{\rho}{m_{\rm{p}}}(1+z_{\rm{snap}})^3
\end{equation}
where $e_{\rm{ab}}$ is the electron abundance, $X_{\rm{H}}$ is the the hydrogen mass ratio, $\rho$ is density, $m_{\rm{p}}$ is the mass of proton and $z_{\rm{snap}}$ is the redshift of the corresponding snapshot. $(1+z_{\rm{snap}})^3$ converts the electron number density in the simulation comoving unit to the proper unit. 
If the distance from the line segment to the nearest gas cell exceeds the cell radius, $r_{\rm{cell}}$, we treat the segment as not intersecting any gas cell and assign it an electron density of zero. The cell radius is calculated as $r_{\rm{cell}}=(3V_{\rm{cell}}/4\pi)^{1/3}$, where $V_{\rm{cell}}$ is the volume of the gas cell, assuming a spherical shape for the cell.
We also exclude star-forming gas from our calculations, as the determination of electron abundance relies on the ``effective" temperature from the equation of state, which is not accurate for star-forming regions. 
Finally the DM of the given sightline is derived by 
\begin{equation}
    \dm=\sum n_{\rm{e,prop}}l_{\rm{prop}}/(1+z_{\rm{snap}})
\end{equation}
where $l_{\rm{prop}}$ is the proper length of each line segment, converted from the comoving unit by $l_{\rm{prop}}=l/(1+z_{\rm{snap}})$. $l$ is the line segment length calculated in the comoving coordinate and the summary is over all the line segments along the sightline. 

\begin{figure}
    \centering
	\includegraphics[width=\columnwidth]{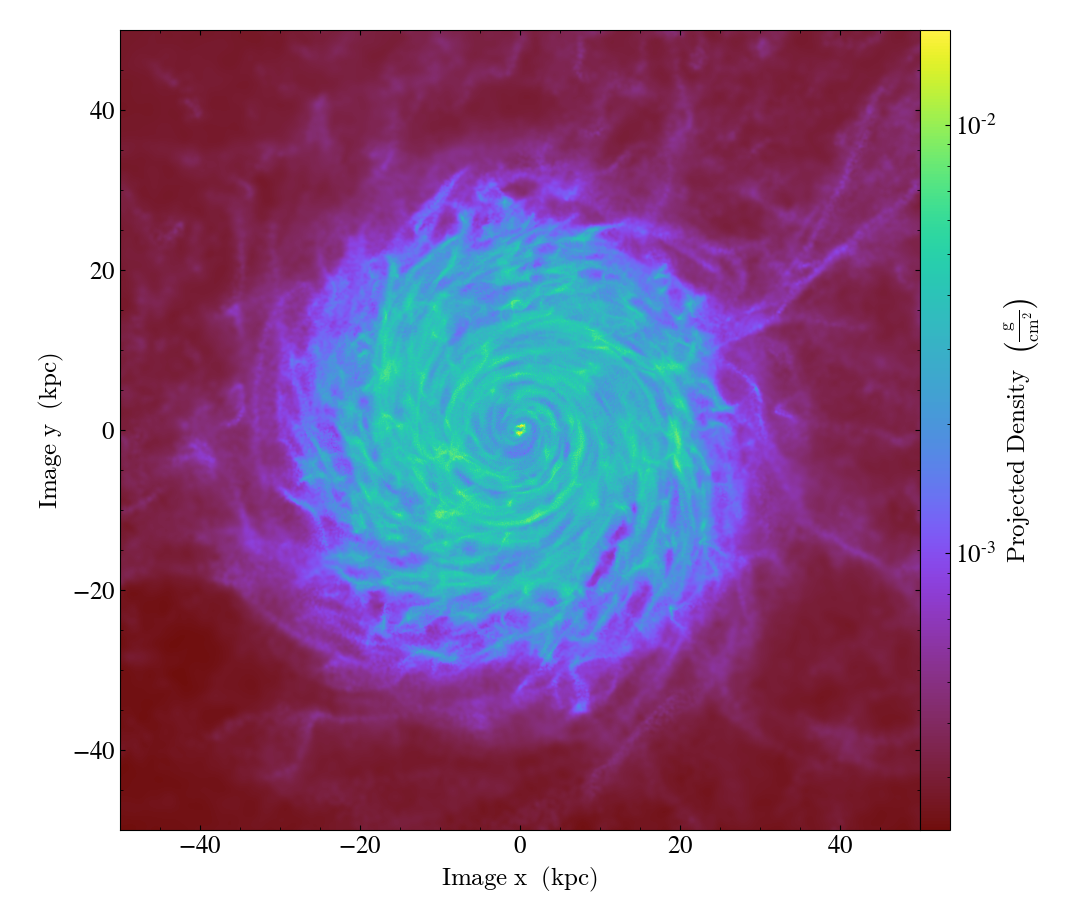}
    \caption{The face-on projection of the MW gas density from the $z=0$ snapshot of the 37\_11 simulation run. This is for illustrative purposes only, as most of the gas shown here is from disk, for which we use the NE2001 model. We use this specific simulation run to model the halo gas component.}
    \label{fig:hestia_mw}
\end{figure}

\begin{figure}
    \centering
	\includegraphics[width=\columnwidth]{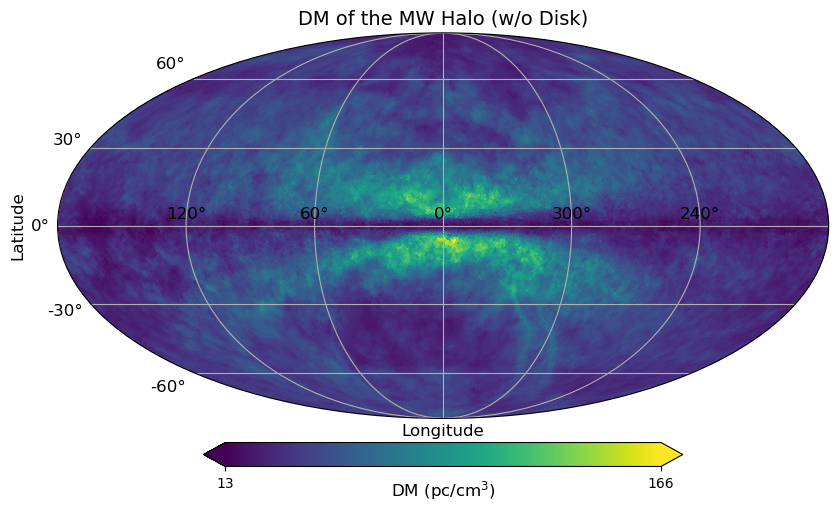}
        \includegraphics[width=\columnwidth]{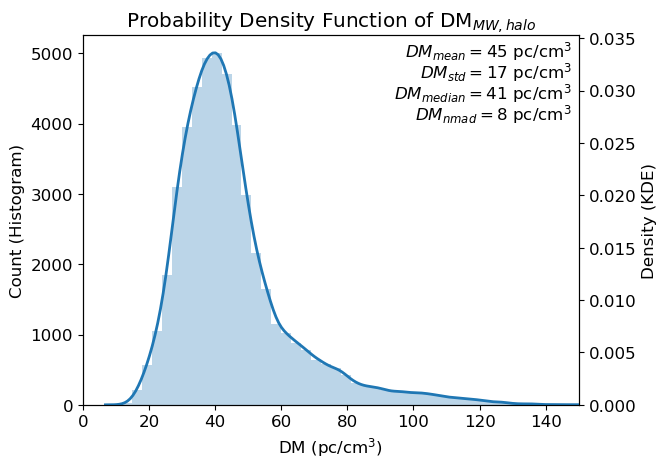}
        \includegraphics[width=\columnwidth]{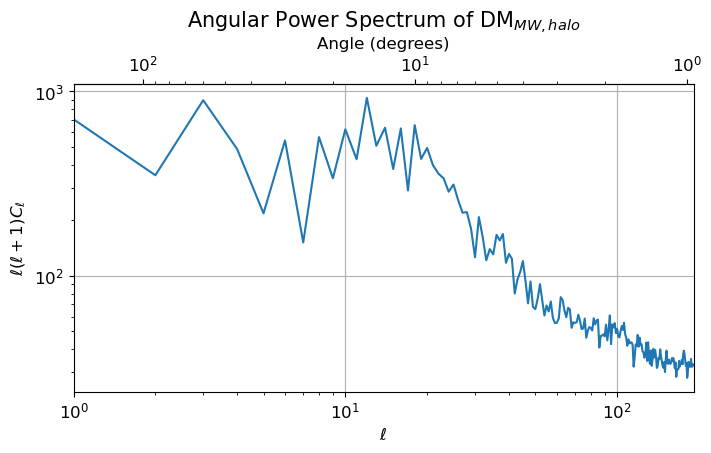}
    \caption{(a) The projected DM model of the MW halo calculated from the \hestia{} simulation displayed in a HEALPix map. The colorbar represents the DM value (instead of $\log\dm$).
    (b) The distribution of the DM contributed by the MW halo of our model. The mean DM of the MW halo contribution is $\langle\dmmwhalo\rangle=45\dmunits$, the standard deviation of the distribution is $\sigma_{\rm{MW,halo}}=18\dmunits$, the median is $17\dmunits$ and the normalized median absolute deviation (NMAD) is $\sigma_{\rm{NMAD}}=4\dmunits$. (c) The angular power spectrum of the MW halo DM map. } 
    \label{fig:dm_mwhalo}
\end{figure}

We use snapshot 127 from the 37\_11 simulation run, which corresponds to redshift 0. The projected gas density of the MW in this snapshot is shown from the face-on perspective in Fig.\ref{fig:hestia_mw}. Before computing the DM model of the corresponding halo, we first determine the Sun's position in this snapshot based on the following criteria: (i) The Sun resides within the MW disk and is 8.2 kpc from the Galactic Center \citep{Leung2023}; (ii) The Galactic longitude of M31 matches the observed value of $l_{\rm{M31}}=121.17^\circ$ (from The NASA/IPAC extragalactic database, \citet{Helou1991}). 
The MW halo DM is then computed within the \rvir{} sphere (see Table \ref{tab:mw_m31_info} for the value) but excluding the region inside the cylinder defined by the ISM DM model computed from NE2001. The resulting DM model of the MW halo, derived from the simulation, along with the DM distribution and the angular power spectrum of this map, is presented in as an all-sky Mollweide projection, in Galactic coordinates, in Figure \ref{fig:dm_mwhalo}. 

In the simulation, the DM of the MW halo ranges from 13 \dmunits{} to 166 \dmunits{}. The middle panel of Figure~\ref{fig:dm_mwhalo} shows the distribution and mean value of our halo, with a mean DM of 45 \dmunits{} and a standard deviation of 18 \dmunits, consistent with the model presented in \citet{Prochaska2019} which predicts $\dmmwhalo\approx50$ \dmunits{} and $\sigma_{\dm}\approx20$ \dmunits.

The projected map in Figure~\ref{fig:dm_mwhalo}a shows a clear excess in a $\sim 10-20$ degree region around the Galactic center, as well as an X-shaped structure extending outwards. 
This might be related to the X-shape structure as observed in the stellar distribution of the real Milky Way \citep{Ness_and_Lang2016}, which is believed to arise due to dynamical instabilities in the disk \citep{Debattista2006,Martinez-Valpuesta2006}.
However, we caution that the veracity of such a structure in the MW gas is yet to be confirmed, so its prediction in Figure~\ref{fig:dm_mwhalo}a should be regarded as hypothetical at this point.
There is also the possibility that this structure is instead related to the Galactic X-ray bubbles detected in eROSITA \citep{Predehl2021}, but a detailed investigation of this is beyond the scope of this paper.
The angular power spectrum of this halo DM distribution (Figure~\ref{fig:dm_mwhalo}) shows that most of the fluctuations in the DM is at $l \lesssim 20$, corresponding to $\gtrsim 10$ degree angular scales on the sky. The fluctuations fall off rapidly with smaller scales and become small at $l\gtrsim 100$, i.e. scales below a degree on the sky.
This power spectrum might be useful, in the future, to construct parametric models of the real MW halo gas distribution that do not rely on specific simulation predictions.

\section{The Local Group}\label{sec:lg}

We now construct a all-sky projection DM model for the Local Group IGrM, also based on the 37\_11 \hestia{} simulation run, centered on the simulated Galactic center. This model will include the intra-group medium (IGrM), as well as the M31 (Andromeda) galaxy and the Magellanic clouds.

\subsection{Intra-group Medium}\label{subsec:igrm}

\begin{figure}
    \centering
	\includegraphics[width=\columnwidth]{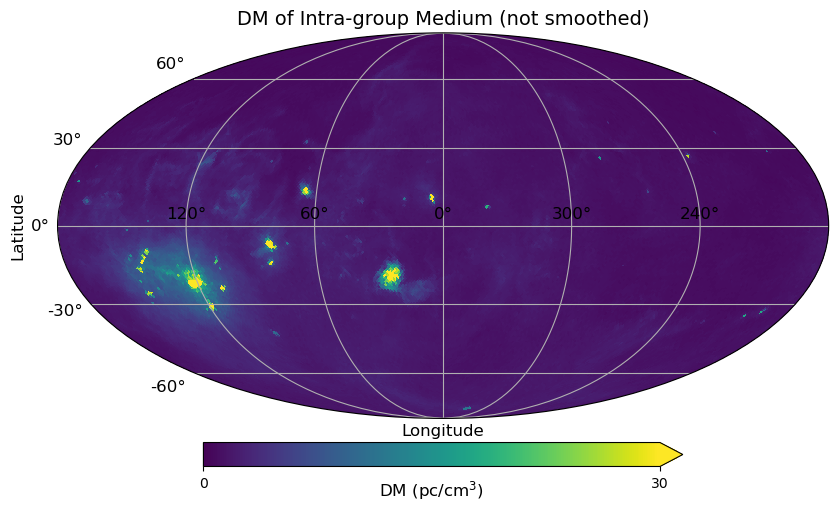}
    \includegraphics[width=\columnwidth]{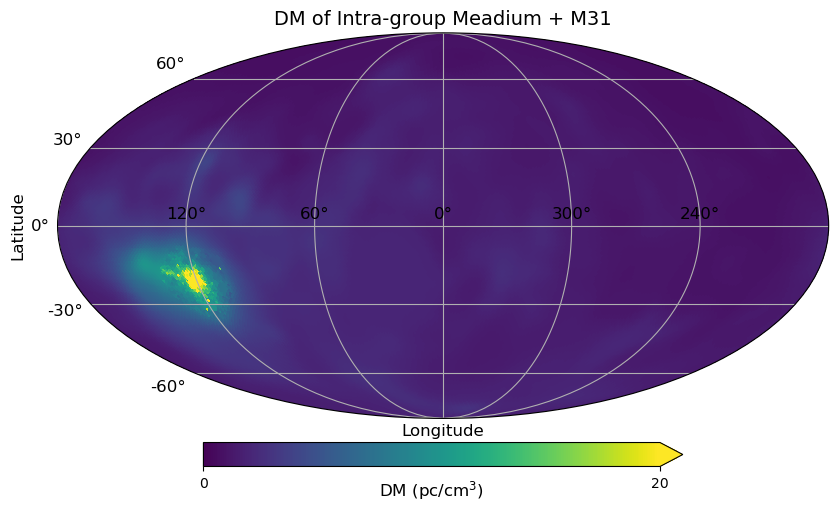}
    \caption{(a) The projected DM model of the IGrM region displayed in a HEALPix map. (b) The DM model of the IGrM after removing the contribution from artificial satellite galaxies that were not constrained to match observations.}
    \label{fig:dm_lgigrm}
\end{figure}

As described in Section \ref{subsec:mwhalo}, the position of the Sun within our model is chosen to ensure that the simulated M31 is at the correct Galactic longitude. To also match the Galactic latitude of the simulated M31 with the real value, we spherically rotate the $r>\rvir$ portion of \hestia, adjusting M31's position along the longitude until the latitude aligns correctly ($b_{\rm{M31}}=-21.57^\circ$; \citet{Helou1991}). 
We then compute the DM of the simulated $r>\rvir$ volume similar to the MW halo. The resulting HEALPix map of the simulated Local Group is shown in Fig.\ref{fig:dm_lgigrm}(a). 
In this figure, the most obvious feature is an extremely high DM region around $l_{\rm{M31}}=121.17^\circ$, $b_{\rm{M31}}=-21.57^\circ$, which can be attributed to M31. However, there are also contributions from satellite galaxies in the simulated Local Group, which we do not expect to correspond to real satellites in the Local Group because \hestia{} is not designed to accurately reproduce objects with masses below MW and M31. We refer to these as `artificial' dwarf galaxies that need to be removed in order to yield a realistic model of our Local Group.

Apart from the simulated M31 and artificial satellites, we find only a negligible ($\dm \sim 2\,\dmunits$) level of dispersion in the Local Group beyond the MW halo.  
In other words, there is no true IGrM permeating the Local Group. This reflects the fact that the Local Group is not a true galaxy group in the sense usually adopted in extra-galactic astronomy, i.e.\ a virialized system dominated by a central galaxy. Rather, the Local Group is best thought of as a merging system of two halos --- the MW and M31 ---  whose respective CGM have yet to interact meaningfully. 

To remove the DM contribution from the artificial satellites, we adopted the following procedure: (i) We replaced the $R<\rvir$ region of each artificial galaxy with the DM value averaged across their surrounding spherical shell at $\rvir<R<1.1\rvir$  (ii) We smoothed the map using a Gaussian kernel of 3 degrees. (iii) The $R<\rvir$ region of M31 from the original map was then embedded into the smoothed map. (iv) Bright points corresponding to artificial satellites within the M31 halo were removed similarly to step (i). After this correction, the resulting map is shown in Fig.\ref{fig:dm_lgigrm}(b). The step (i) removes $\sim 2.6$ \dmunits{} per deg$^2$ and totally $\sim 2317$ deg$^2$ of contamination from artificial galaxies. This approach can largely preserves the local group environment, which is primarily influenced by the MW and M31.

\subsection{Dwarf Galaxies}\label{subsec:dwarfs}
\begin{table*}
	\centering
	\caption{The stellar masses of the LMC, SMC, and M33, along with their distances to their respective host galaxies (LMC and SMC relative to the MW, and M33 relative to M31), as derived from observations. Those of their simulated counterparts are also listed for comparison.}
	\label{tab:counterparts}
	\begin{tabular}{lccccc} 
		\hline
		  & Observed \mstar & Observed Distance & Counterpart \mstar & Counterpart Distance & Reference \\
		\hline
		LMC & $2.7\times10^9\msun$ & 50 kpc & $1.6\times10^9\msun$ & 91 kpc & \mstar: \citet{Besla2015}; Distance: \citet{Pietrzynski2019} \\
		SMC & $3.1\times10^8\msun$ & 62 kpc & $2.6\times10^8\msun$ & 158 kpc & \mstar: \citet{Besla2015}; Distance: \citet{Graczyk2020} \\
        M33 & $2.3\times10^9\msun$ & 220 kpc & $1.3\times10^8\msun$ & 213 kpc & \mstar: NEDLVS; Distance: \citet{Breuval2023} \\
		\hline
	\end{tabular}
\end{table*}

\begin{figure*}
    \centering
    \begin{minipage}[t]{0.33\textwidth}
        \centering
        \includegraphics[width=\textwidth]{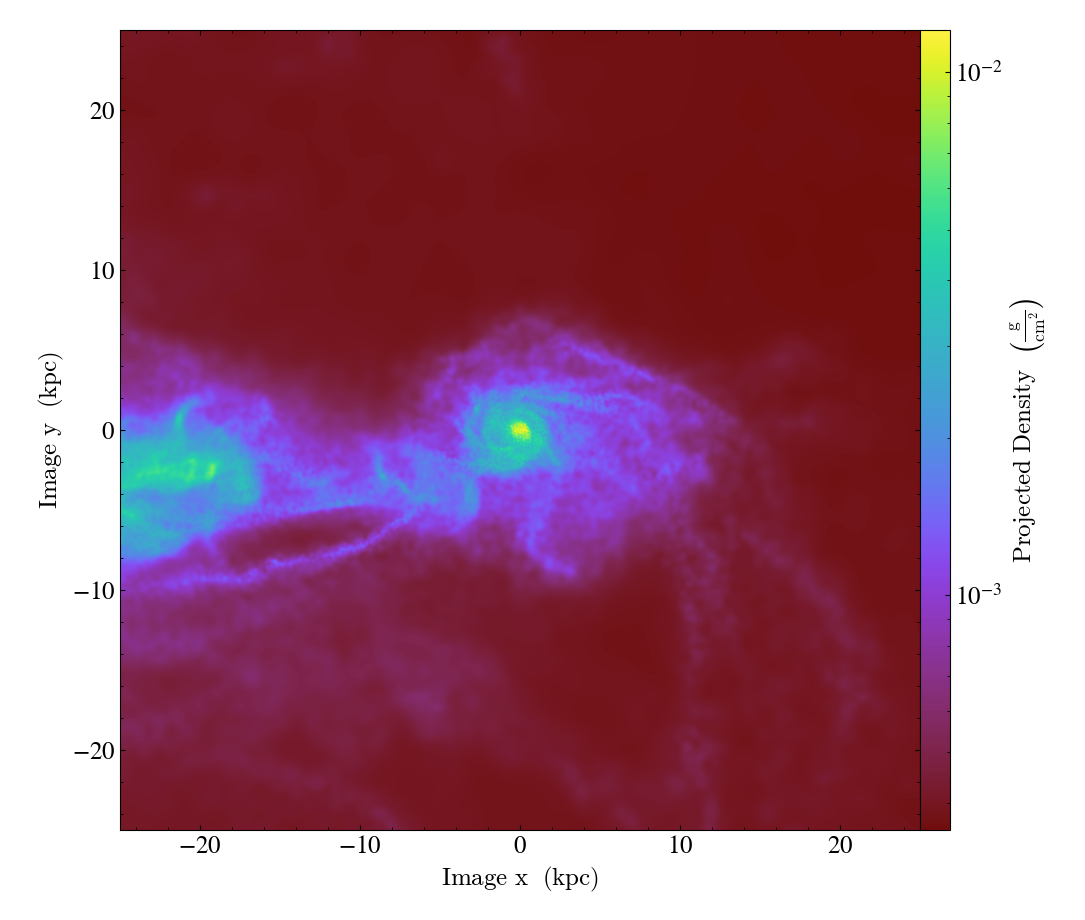}
    \end{minipage}
    \hfill
    \begin{minipage}[t]{0.33\textwidth}
        \centering
        \includegraphics[width=\textwidth]{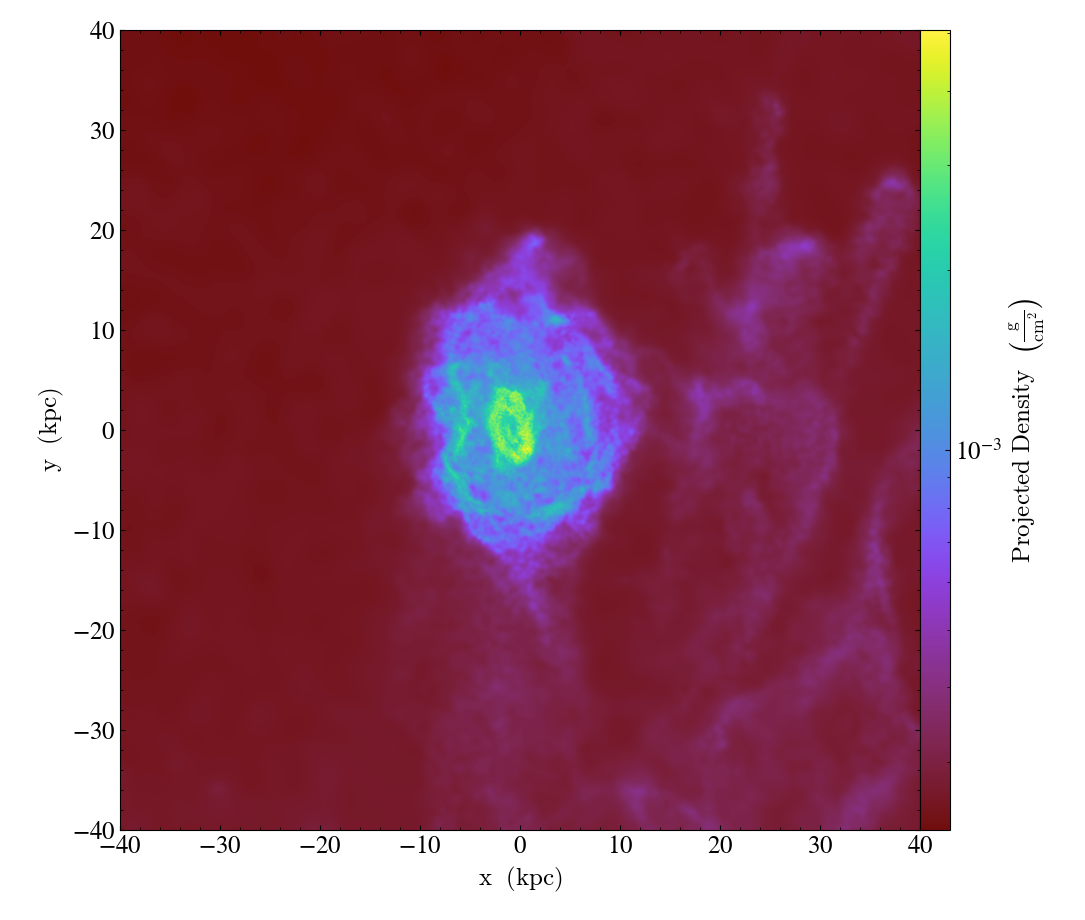}
    \end{minipage}
    \hfill
    \begin{minipage}[t]{0.33\textwidth}
        \centering
        \includegraphics[width=\textwidth]{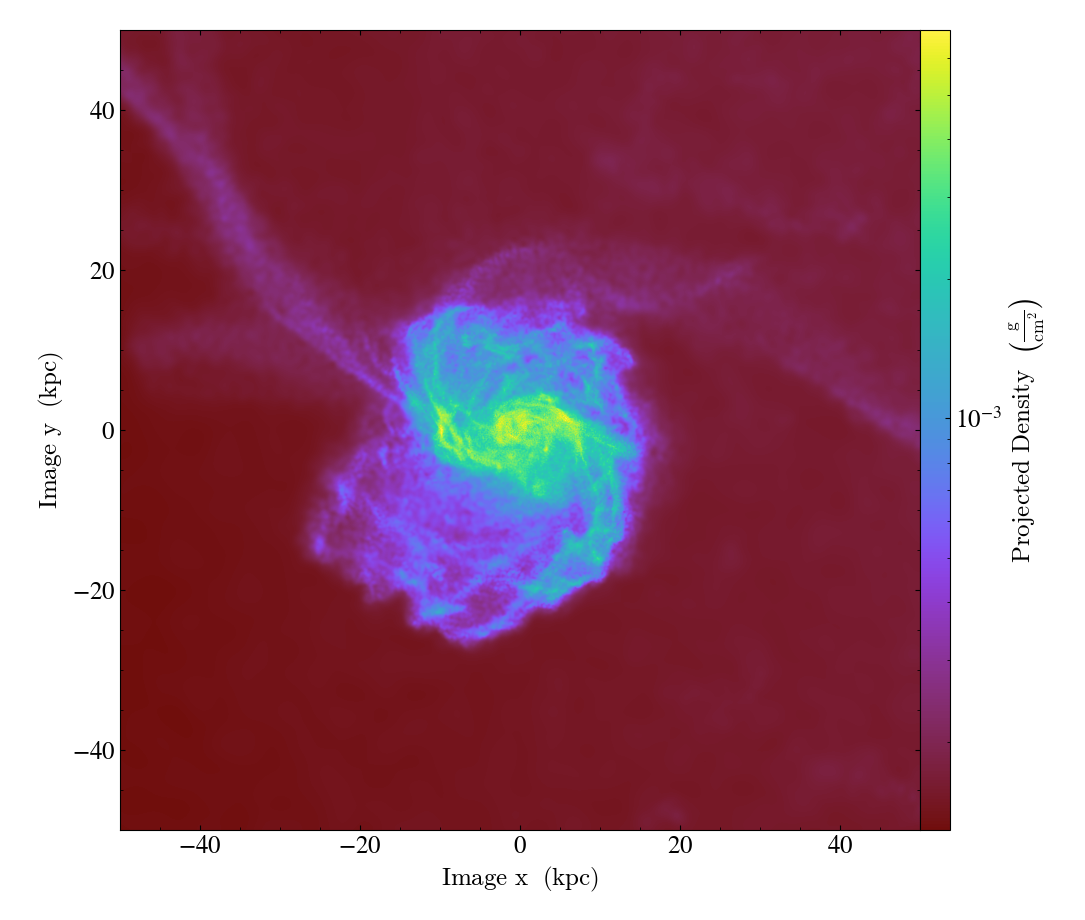}
    \end{minipage}
    \caption{Projected gas density of the simulated counterparts of LMC (left), SMC (middle) and M33 (right). These maps are centered on their own centres-of-mass}
    \label{fig:counterparts}
\end{figure*}

With the artificial satellite galaxies removed, we now manually introduce into our model the three largest dwarf galaxies within the Local Group: Large Magellanic Cloud (LMC), Small Magellanic Cloud (SMC), and Triangulum Galaxy (M33). 
Because we do not expect the \hestia{} volumes to reproduce their proper locations within the Local Group, we proceed by searching for their simulated counterparts and inserting them into their true locations on the sky.
These simulated counterparts were selected based on the following criteria: (i) The stellar mass of the simulated counterparts should be within 25\% from the observed stellar mass of the respective dwarf galaxies; (ii) The counterparts must be satellite galaxies of either MW or M31 in the simulation; (iii) From the galaxies that meet criteria (i) and (ii), we select the simulated counterparts of the three dwarf galaxies whose distances to their host galaxy are the closest matches to the observed distances of the corresponding dwarf galaxies and their hosts. Due to the limited number of satellite galaxies around MW and M31 in the 37\_11 simulation alone, we made this selection across all three highest-resolution runs within the \hestia{} suite. It is important to note that since \mhalo{} and \rvir{} of MW and M31 are similar across these simulations, their environments do not differ dramatically. Therefore, in principle, the names of MW and M31 can be interchanged in each simulation, allowing us to search for counterparts among the satellite galaxies of both simulated MW and M31 for all three dwarf galaxies. The stellar masses of the dwarf galaxies and their distances to their host galaxies, along with those of their simulated counterparts, are listed in Table \ref{tab:counterparts}. The projected gas density of the simulated counterparts is shown in Fig.\ref{fig:counterparts}, which reveals their gas exchange with their host galaxies.

\begin{figure}
    \centering
	\includegraphics[width=\columnwidth]{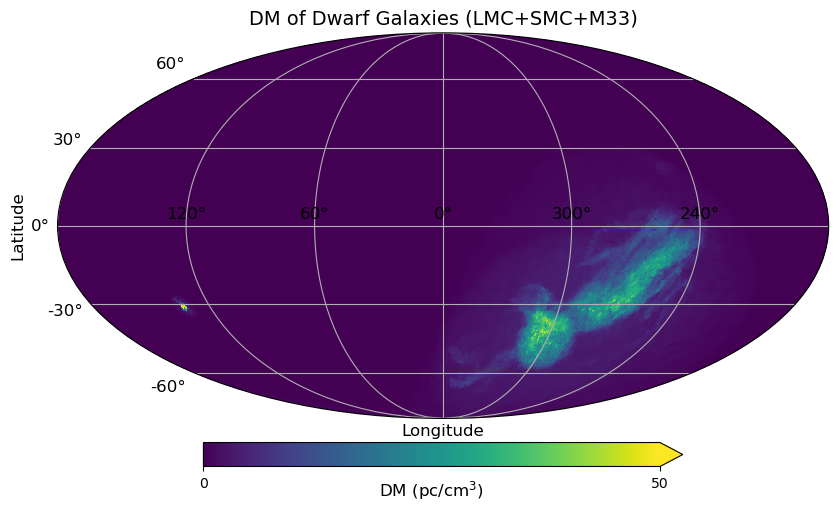}
    \caption{The projected DM model of the simulated counterparts of LMC, SMC and M33 displayed in a HEALPix map. The model ensures the distances from simulated counterparts to ``the Sun" and their 
    inclination match the observation. Only the DM contributed by the CGM within one \rvir{} is included.}
    \label{fig:dm_lgdwarfs}
\end{figure}

To calculate the DM model for these counterparts, the position of 'the Sun' is selected along the line of sight from the simulated counterparts to their host galaxies, with the distance from 'the Sun' to the simulated counterparts set to match their observed distance. Then the simulated counterparts are rotated to match their observed inclination ($i_{\rm{LMC}}=35^\circ$, $i_{\rm{SMC}}=40^\circ$, $i_{\rm{M33}}=57^\circ$; \citealt{Bekki2009,Kourkchi2020}). Only the segments within the galaxy halos ($<\rvir$) are included in the DM model calculation. Finally, the projected DM model of these simulated Local Group satellites is computed and presented in Fig.\ref{fig:dm_lgdwarfs}. 

It is clear from Figure~\ref{fig:dm_lgdwarfs} that the projected DM from the LMC and SMC extends across enormous ($\sim 10$ deg) angular scales on the sky, with contributions of up to $\dm \sim 40\,\dmunits$.
This reflects their small distance ($D\sim 50-60\,$kpc) from the MW, and in fact their CGM overlaps with the MW disk. 
Given the inclination of the LMC, our simulation predicts that its projected DM signal would be in a planar configuration from the perspective of the MW observer, while the SMC is somewhat rounder.
M33, being M31's satellite located at a larger distance from the MW, has a much smaller projected size. 

\section{The Local Universe}\label{sec:localU}

The DM model for the remaining part of the Local Universe beyond the simulated volume of \hestia{} (comoving distance 3.4 \Mpc{} $<D<120$ \Mpc) consists of contributions from the IGM and galaxy halos. The IGM component is derived from the HAmiltonian Monte carlo reconstruction of the Local EnvironmenT (\hamlet) reconstructed density fields of the Local Universe \citep{Valade2024}, while the contribution from individual galaxy halos is calculated using data from the 2MASS Galaxy Group Catalog and the NEDLVS Catalog. Both components of the projected model are computed at different distance layers, spanning from 4 Mpc to 120 Mpc, with thickness of 8 Mpc per shell.
The outermost distance for our Local Universe model is set approximately by the outer limit where galaxy redshift data is available for the full sky, while ensuring that the structures associated with Coma Cluster ($D \sim 100-110\,\mathrm{Mpc}$) are fully covered.

\subsection{Intergalactic Medium}\label{subsec:IGM}

The matter density field in the Local Universe is reconstructed with the \hamlet{} code \citep{Valade2022} applied to the Cosmicflows-4 \citep[CF4][]{Tully2023} catalog containing about 56\,000 measurements of peculiar velocities and is relatively dense up to $D<120\,\Mpc$ (except in the Zone of Avoidance). The \hamlet{} method is a field-level inference of both density and velocity field made possible by a powerful Hamiltonian Monte Carlo technique. The fully Bayesian approach tackles the difficulties arising from the complexity of peculiar velocity data \citep[{\it e.g.} the Log-normal bias][]{Hoffman2021} which may curb the more traditional Wiener Filter/Constrained Simulation pipeline \citep{Valade2023} as used for the setting up of the \hestia{} simulations. Yet, both approaches are limited by their usage of the linear theory which needs to bind the density fields to the velocity measurements, yielding results accurate to a scale of $> 5\,\Mpc/h$. 

The density field used in this work is the mean of 1000 possible density fields (Monte Carlo states) projected on a periodic box of $1 \,{\rm Gpc}/h$ of edge with $256$ nodes per side, i.e a spatial resolution of $3.9\,\Mpc/h$ \citep{Valade2024}. Notable reconstructed features of the Local Universe within $D<120\,\Mpc$ include the clusters Virgo, Coma, Norma and Perseus as well as voids such as the Local Void, the Sculptor Void or Hercules Void.

Similar to the method described in Section \ref{subsec:mwhalo}, each sightline is uniformly divided into 1000 segments. We calculate the DM for each segment and sum these to obtain the total DM. The DM can be derived from the \hamlet{} matter over-density field using 
\begin{equation}
    \text{DM}_{\text{igm}} = \bar{n}_e^{\text{igm}} \sum_s \left( 1 + \delta_{m,s}^{\text{sm}} \right) l_s \left( 1 + z_s \right)^{-1}
\end{equation}
where $\delta_{m,s}^{\text{sm}}$ is the smoothed matter over-density field of line segment s, $l_s$ and $z_s$ are the length and the redshift of the line segment s, respectively. $\bar{n}_e^{\text{igm}}$ is the mean cosmic electron density at redshift 0 and is defined as:
\begin{equation}
    \bar{n}_e^{\text{igm}} = f_{\text{igm}} \Omega_b \bar{\rho}_c \left[ \frac{m_{\text{He}}(1 - Y) + 2Y m_{\text{H}}}{m_{\text{H}} m_{\text{He}}} \right]
\end{equation}
where $\Omega_{\rm{b}}=0.048$ is the baryon fraction of the critical density, $\bar{\rho}_c$ is the critical density of Local Universe (here we simply assume redshift 0), $m_{\rm{H}}$ and $m_{\rm{He}}$ are the atomic masses of hydrogen and helium, respectively, $Y=0.243$ is the cosmic mass fraction of helium, and $\figm$ is the fraction of cosmic baryons residing in the diffuse IGM.
The true value of $\figm$ is a topic of active investigation \citep{Khrykin2024b,Connor2024}, but ranges from $0.6\lesssim \figm \lesssim 0.85$ in cosmological hydrodynamical simulations \citep{Jaroszynski2019,Khrykin2024a,Walker2024}.
We therefore set $\figm$ as a free parameter that can be specified by the user, but will use $\figm=0.8$ as a default placeholder value. 

\begin{figure}
    \centering
	\includegraphics[width=\columnwidth]{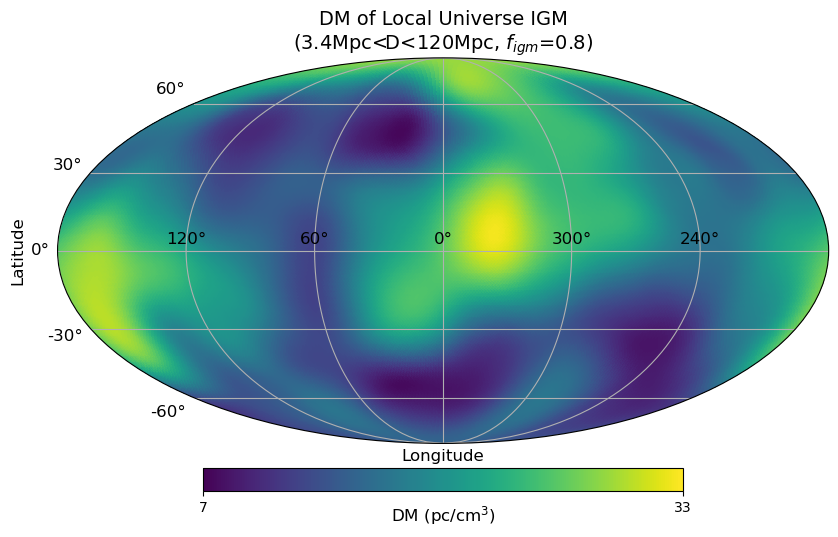}
    \caption{The projected DM model of the Local Universe IGM from 3.4 Mpc to 120 Mpc is displayed on a HEALPix map, with the positions of major structures such as the Virgo Cluster, Norma Cluster, Coma Cluster, and Perseus-Pisces Supercluster labeled. We adopt $\figm=0.8$ in this map.}
    \label{fig:dm_igm}
\end{figure}

We apply an 8 $\Mpc\,h^{-1}$ Gaussian smoothing kernel to the matter density field and linearly interpolate the smoothed field. The smoothing scale is chosen to be larger than the Nyquist wavelength, which is twice the grid size, and to ensure that $1+\delta_m^{\text{sm}}$ remains positive throughout the box. Finally, we compute the DM IGM model with the interpolated matter density field. The projected \dmigm{} out to 120 Mpc is shown in Fig.\ref{fig:dm_igm}, adopting $\figm=0.8$.
At the center of this projected map, we see that the \hamlet-based model has captured the structures surrounding the Norma galaxy cluster (Abell 3627), a massive cluster ($\mhalo \sim 10^{15}\,\msun$) that has historically been under-studied because it lies within the Milky Way's Zone of Avoidance.
However, the largest DM contributions ($\dmigm \sim 30\,\dmunits$) arise from the large-scale structure in the vicinity of the Coma cluster (Abell 1656), which does not show up well in this projection because it is nearly coincident with the Northern Galactic Pole. The excess DM extends across $\sim 10$ deg scales, well beyond the $\rvir\sim 2.7\,\Mpc$ virial radius of Coma cluster \citep{Kubo2007}. The structures associated with Virgo cluster (Abell 1689) are also visible at a high Galactic latitude.
Conversely, one also sees large extended regions with low dispersion ($\dmigm \sim 4-5\,\dmunits$), several times smaller than the mean cosmic value expected at that distance. These represent sightlines that traverse multiple cosmic voids in the Local Universe.

The publicly available map of our Local Universe \dmigm{} model is comprised of 13 concentric distance shells integrated out to distances ranging from 4 Mpc to 120 Mpc, in 8 Mpc intervals. The \dmigm{} value at arbitrary distances within $D<120$\,Mpc can be derived by linearly interpolating between these layers. 

\subsection{Galaxy Halos and Clusters}\label{subsec:halos}

We therefore compute a separate DM model for foreground galaxy halos and clusters in the Local Universe, \dmhalos{}, based on the NEDLVS catalog \citep{CookDO2023} in conjunction with the 2MASS Galaxy Group Catalog \citep[Tully15 hereafter]{Tully2015}. 
While some of the FRB literature uses the nomenclature \dmigm{} to refer to all diffuse gas outside of galaxies, the halo gas (CGM/IGrM or ICM depending on mass) and IGM are quite physically different and are expected to have different baryonic fractions \citep[e.g.][]{Khrykin2024a, Khrykin2024b}.
In our notation, we further make the distinction between plural $\dmhalos \equiv \sum_i \mathrm{DM}_{\mathrm{halo},i}/(1+z_i)$ and the singular $\dmhalo$, where the former refers all contributions along a sightline while the latter is the contribution from a individual halo.

Note that while the previous components of our DM model were computed as HEALPix maps with a pixelisation of $\sim$1 degree, 
the \dmhalo{} can vary significantly even on arcsecond angular scales depending on the intersected foreground halos, and where in the halos they are intersected. 
Attempting to construct a useful HEALPix map of the all-sky extragalactic halo contribution would therefore lead to huge file sizes that are inconvenient to share.
Instead, our model computes the halo contribution on-the-fly from the galaxy and group catalogs, based on the desired sky coordinates.

For a given coordinate, we select NEDLVS galaxies within a transverse separation of 1 Mpc, excluding those with insecure redshifts (\texttt{z\_qual=1}) and galaxies with $\mstar<10^{13}\msun$ (that will be modeled separately; below). We then calculate their \mhalo{} using the \citet{Moster2013} mean stellar-to-halo mass relation. 

For galaxy halos with $\mhalo<10^{13.5}\msun$,
we model the distribution of hot, virialized gas using the Modified NFW (mNFW) model \citep{Prochaska2019}, where the baryon density profile is given by:
\begin{equation}
    \rho_b(r) = \fgas\frac{\Omega_b}{\Omega_m} \frac{\rho_0(\mhalo)}{y^{1-\alpha}(y_0 + y)^{2+\alpha}},
\end{equation}
$f_{\text{gas}}$ represents the fraction of cosmic baryons that reside in the CGM of each individual galactic halo, compared to the amount that would be present if the cosmic baryon fraction were achieved within the halo. $\rho_0(\mhalo)$ is the central density of the halo and is a function of \mhalo. $y \equiv c(r/r_{200})$ and $c$ is the concentration parameter. $y_0$ and $\alpha$ are the parameters of the mNFW model.
In our model, we assume the $\fgas=0.25$, $y_0=2$, $\alpha=2$ and the radial extent of the gas profile for each halo $\rmax=1\rvir$ by default.

As for clusters with $\mhalo>10^{13.5}\msun$, we employ the intracluster medium (ICM) model from \citet{Vikhlinin2006}, setting $\fgas=0.8$ by default.
In both models, the maximum radius of the gas profile is set to one \rvir. Both the mNFW and ICM models are available in the FRB repository \citep{FRBsoft2023}. With the gas profile and parameters defined, the DM model for galaxy halos and clusters can be easily computed using the \texttt{Ne\_Rperp} function from the repository.

\begin{figure}
    \centering
	\includegraphics[width=\columnwidth]{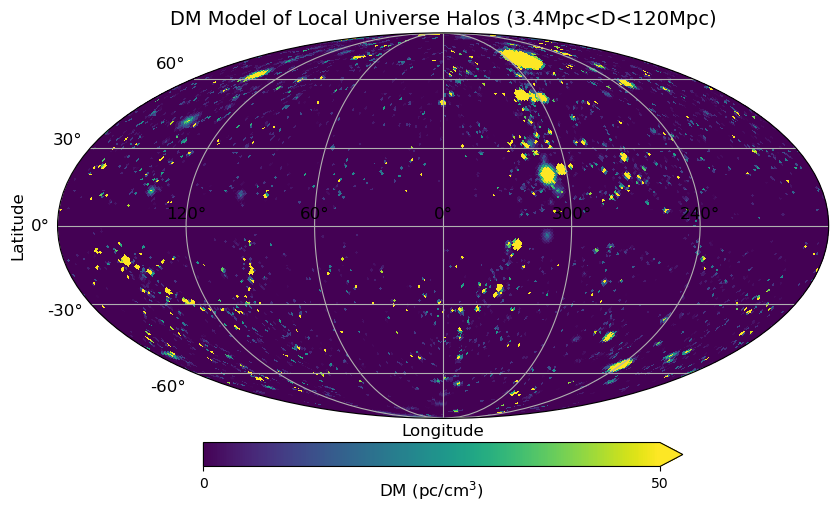}
        \includegraphics[width=\columnwidth]{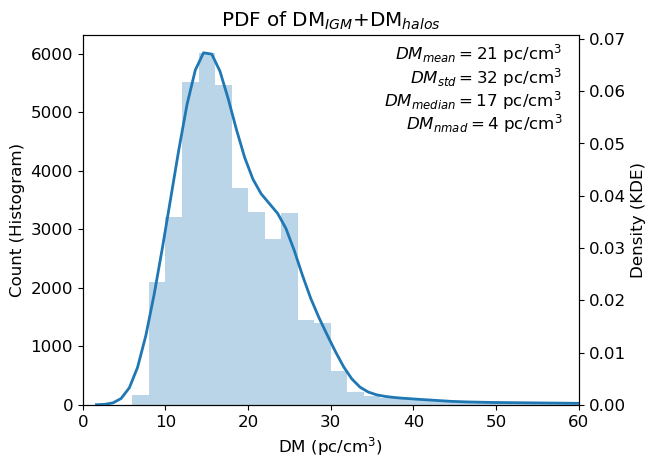}
    \caption{(a) The projected DM model of galaxy halos and clusters from 3.4 Mpc to 120 Mpc is displayed on a HEALPix map. In this map, we apply the mNFW model for galaxy halos with $\fgas=0.25$, $y_0=2$ and $\alpha=2$; and the ICM model for clusters with $\fgas=0.8$. Note that we have truncated the map intensity scale at $\dmhalo = 50\,\dmunits$ for visual clarity --- the \dmhalo{} values are much larger than this in the inner regions of massive halos. (b) The PDF of the projected DM model for the IGM and halos from 3.4 Mpc to 120 Mpc, with a mean DM contribution of $\langle\dmigm+\dmhalos\rangle=21\dmunits$, a standard deviation of 32 \dmunits, a median of 17 \dmunits, and an NMAD of 4 \dmunits.}
    \label{fig:dm_halos}
\end{figure}

For a given sightline, we also select galaxy groups and clusters from the Tully15 catalog with impact parameters within 5 Mpc. These groups are matched with the corresponding NEDLVS galaxies and clusters. For the matched groups and clusters, we remove the DM contributions derived from the NEDLVS catalog and recalculate the DM using the \mhalo {}from the Tully15 catalog. For visualization purposes, we evaluate our $3.4\,\mathrm{Mpc} < D < 120\,\mathrm{Mpc}$ halo DM model at the pixel centers of the HEALpix projection used in the MW and Local Group components, and show the sky distribution in Fig.\ref{fig:dm_halos}.

Our software package includes the relevant NEDLVS and Tully15 catalogs and calculates the DM contribution from Local Universe halos along arbitrary sightlines. The model can also accommodate \fgas{} as an arbitrary function of \mhalo{} as specified by the user, in which case we apply the mNFW model to both galaxy halos and clusters.

\begin{table*}
	\centering
	\caption{Summary of the DM contributions from various components of the model, including the MW halo (with and without contributions from the LMC and SMC), the Local Group IGrM, the IGM, the halos, and the model combining all the aforementioned components (with and without the NE2001 model). For each component, the table presents the distance range, the mean DM, the standard deviation, the median DM, the NMAD, and the ZoA, which is indicated as the Galactic latitude range where the model has increased uncertainty.}
	\label{tab:model_summary}
	\begin{tabular}{lcccccc} 
		\hline
		  DM Model Component & Distance & Mean  & Standard Deviation & Median & NMAD & ZoA  \\
                           & (Mpc) & (\dmunits) & (\dmunits) & (\dmunits)  & (\dmunits) & \\
		\hline
            MW halo (no LMC \& SMC) & $D<0.2$ & 45 & 17 & 41 & 8 & / \\
            MW halo (with LMC \& SMC) & $D<0.2$ & 46 & 18  & 42 & 8 & / \\
            MW halo (with LMC \& SMC) + NE2001 & $D<0.2$ & 112 & 138 & 68 & 18 & / \\
            IGrM & $0.2<D<3.4$ & 2 & 3 & 2 & 0.5 & / \\
            IGM & 3.4$<D<$120 & 17  & 5  & 16  & 3  & $|b|<10^\circ$ \\
            Halos & 3.4$<D<$120 & 3 & 31  & 0& 0  & $|b|<10^\circ$ \\
            IGM + Halos & 3.4$<D<$120 & 21  & 32  & 17  & 4 & $|b|<10^\circ$ \\
            Combined (no NE2001) & $D<120$ & 70  & 47 & 63  & 9 & $|b|<10^\circ$ \\
            Combined (with NE2001) & $D<120$ & 96 & 48 & 83  & 15 & $|b|<10^\circ$ \\
		\hline
	\end{tabular}
\end{table*}

\section{Discussion}\label{sec:discussion}

The Zone of Avoidance (ZoA) refers to a region along the Galactic equator where observations of extragalactic objects are challenging due to the dense concentration of dust, gas, and stars in the Milky Way's disk. Since the CosmicFlows-4 dataset which is used to construct the \hamlet{} density field, as well as the NEDLVS and Tully15 Catalogs which are used to compute the \dmhalos{} model, have lower completeness within the ZoA, our DM model exhibits higher uncertainty in this region compared to areas outside it.
In our model, the ZoA is defined as $|b|<10^\circ$. 

\begin{figure*}
    \centering
	\includegraphics[width=0.8\textwidth]{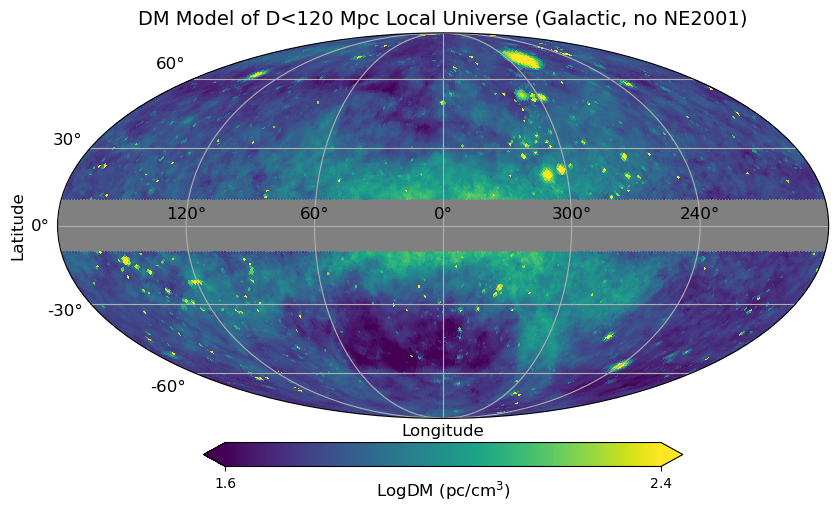}
        \includegraphics[width=0.8\textwidth]{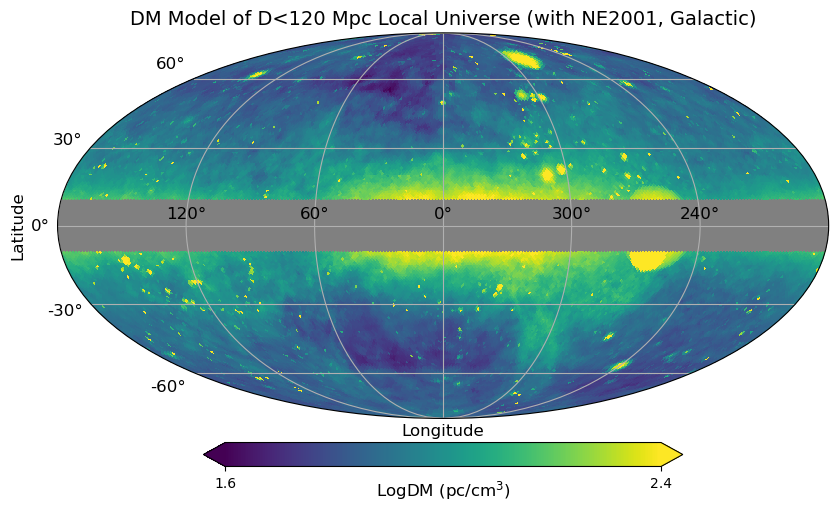}
    \caption{(a) The projected DM model incorporating all components considered in this study, including the MW halo, the Local Group IGrM, the IGM, and galaxy halos within 120 Mpc, but excluding the NE2001 model. (b) The same combined DM model as in (a), but with the NE2001 model included. The grey region in the figure represents the ZoA. The DM values are displayed on a logarithmic scale.}
    \label{fig:dm_combined_galactic}
\end{figure*}

\begin{figure*}
    \centering
	\includegraphics[width=0.8\textwidth]{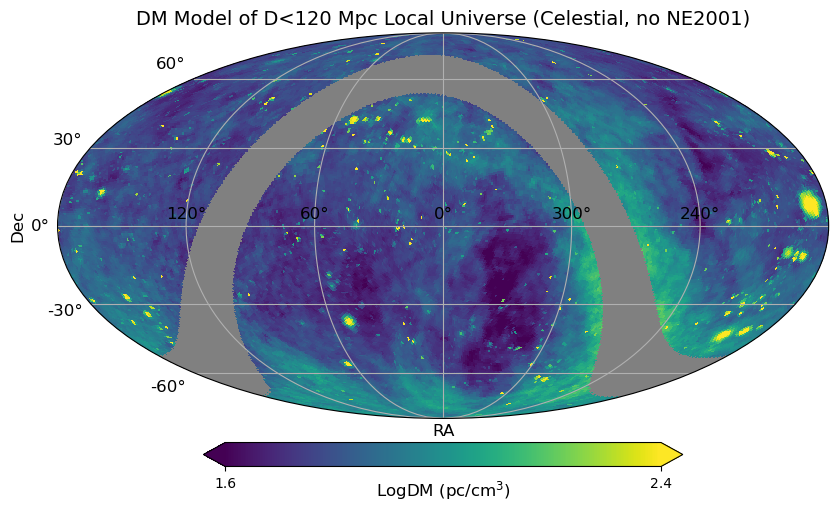}
        \includegraphics[width=0.8\textwidth]{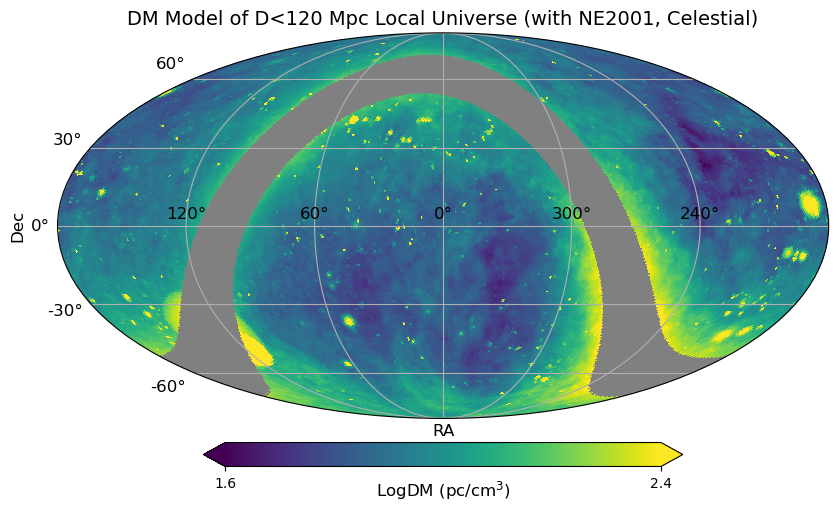}
    \caption{Similar plot as Fig.\ref{fig:dm_combined_celestial} in Celestial Coordinate.}
    \label{fig:dm_combined_celestial}
\end{figure*}

\begin{figure}
    \centering
	\includegraphics[width=\columnwidth]{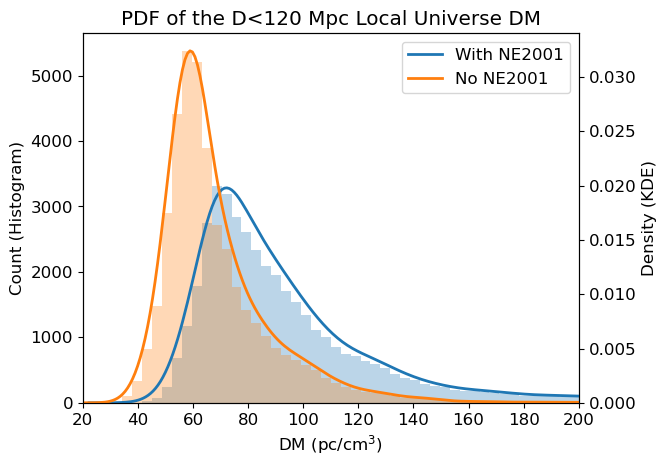}
    \caption{The PDF of the combined $D<120$ Mpc Local Universe DM model with (blue) and without (orange) NE2001 model with the ZoA excluded. For the combined map excluding NE2001, the mean DM value is $\langle\dmlu\rangle=70$ \dmunits, the standard deviation is $\sigma_{\rm{LU}}=47$ \dmunits, the median value is 63 \dmunits, and the NMAD is $9$ \dmunits. For the combined map including NE2001, the mean DM is $\langle\dmlu\rangle=96$ \dmunits, the standard deviation is $\sigma_{\rm{LU}}=48$ \dmunits, the median is 83 \dmunits, and the NMAD is 15 \dmunits.}
    \label{fig:dm_combined_dist}
\end{figure}

The combined \dmlu{} map of all the Local Universe components (the MW disk and halo, the Local Group IGrM, the IGM, and the galaxy halos) is presented in Fig.\ref{fig:dm_combined_galactic} and Fig.\ref{fig:dm_combined_celestial}, in Galactic coordinates and celestial coordinates respectively. The DM contribution from the LMC and the SMC, along with other galaxy halos, are clearly visible on the map. However, the IGM distribution is relatively small and is largely subdominant compared with the contributions from these galaxy halos. Figure \ref{fig:dm_combined_dist} displays the distribution of the combined all-sky DM map of the $D<120$ Mpc Local Universe with and without NE2001, excluding the ZoA. For the combined map excluding NE2001, the mean DM value is $\langle\dmlu\rangle=70$ \dmunits, the standard deviation is $\sigma_{\rm{LU}}=47$ \dmunits, the median value is 63 \dmunits, and the NMAD is $9$ \dmunits. For the combined map including NE2001, the mean DM is $\langle\dmlu\rangle=96$ \dmunits, the standard deviation is $\sigma_{\rm{LU}}=48$ \dmunits, the median is 83 \dmunits, and the NMAD is 15 \dmunits.
For the all-sky projected DM model of each component, the mean value, standard deviation, median value and NMAD is summarized in Table \ref{tab:model_summary}. Note that the median value of the Galactic component (including the MW halo and NE2001) is 68 \dmunits, which is consistent with the empirical estimates of $64\pm20$ \dmunits derived by \citet{Das2021} using 21-cm, UV, EUV, and X-ray data.

The largest DM values ($\dmlu \gtrsim 200\,\dmunits$) seen in Figure~\ref{fig:dm_combined_galactic} arise from the circum-halo gas of discrete halos beyond the Local Group, especially from galaxy clusters. 
However, one also sees large extended regions within $\sim 20$ deg of the MW bulge 
with enhanced DM ($\dmlu \sim 100\,\dmunits$) arising from within the MW halo, with the LMC and SMC contributions spanning comparable angular scales, but at a slightly lower dispersion ($\dm \sim 50\,\dmunits$). 
Conversely, our model predicts extended regions up to 100 degrees across with low total DM ($\dmlu \sim 40-60\,\dmunits$) extending out to $D \sim 100\,\Mpc$, that avoid significant ionized regions in the MW disk, halo, and the Local Group, before traversing multiple cosmic voids out to the distant Universe. 
The most prominent such region is at $30^\circ \lesssim b \lesssim 60^\circ$, at $\pm 10^\circ$ around Galactic longitudes $l \sim 10^\circ$.
Such under-dispersed regions could be part of the explanation for FRBs that fall below the Macquart relation \citep[e.g.,][]{Connor2024}.

\begin{figure}
    \centering
	\includegraphics[width=\columnwidth]{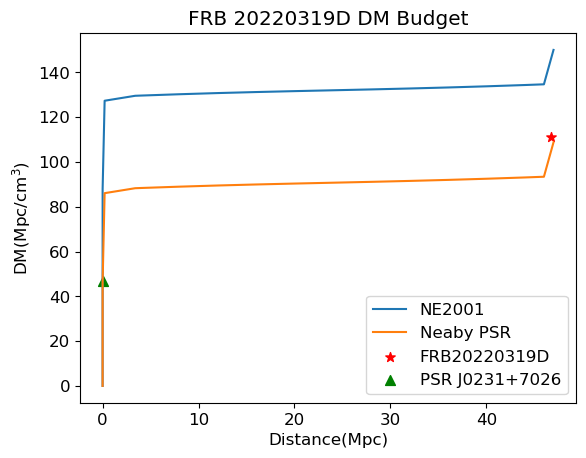}
    \caption{Cumulative DM as a function of distance in the direction of FRB 20220319D. The \dmmwism{} is modeled using two approaches: the NE2001 model (blue) and the DM from the nearby pulsar PSR J0231+7026 (orange), which is separated by 0.9 degrees from the FRB. The remaining DM components are calculated based on the model described in this paper. With the \dmmwism{} value of the pulsar, the toal DM estimated from our model is in rough agreement with that of FRB 20220319D.}
    \label{fig:220319D_dm_budget}
\end{figure}

To test the consistency of our model, we compare with 2 previous studies on constraining the MW DM model. In \citet{Ravi2023}, they reported FRB 20220319D with an observed DM of 110.95\,\dmunits. 
While this FRB is nominally excluded by our ZoA, we went ahead to analyze the DM budget of FRB 20220319D (Figure~\ref{fig:220319D_dm_budget}). We model the \dmmwism{} with two different assumptions: NE2001 model and the DM of the neaby pulsar PSR J0231+7026 which has a 0.9 degrees separation from the FRB (as was done by \citealt{Ravi2023}). Using \texttt{pyhesdm}, we calculate the other components, estimating $\dmmwhalo= 39\, \dmunits$, $\rm{DM_{IGrM}}= 2.2\, \dmunits$, and $\dmigm{} = 5.3\, \dmunits$. For halo contributions, there are no intervening halos and the only significant contribution comes from the host galaxy halo: DM$_{\rm{host,halo}}=15\,\dmunits$. This value is half of the \dmhalos{} computed by \texttt{pyhesdm}, assuming the FRB comes from the center of its host halo. The significant discrepancy between the predicted total DM from NE2001 and the observed DM arises from the large uncertainty in the NE2001 model at low Galactic latitudes, where the FRB is located ($b=9.11^\circ$). 
Once we set \dmmwism{} to that implied by PSR J0231+7026, the predicted total DM from our model is within 5\% of the observed DM.

\begin{figure}
    \centering
	\includegraphics[width=\columnwidth]{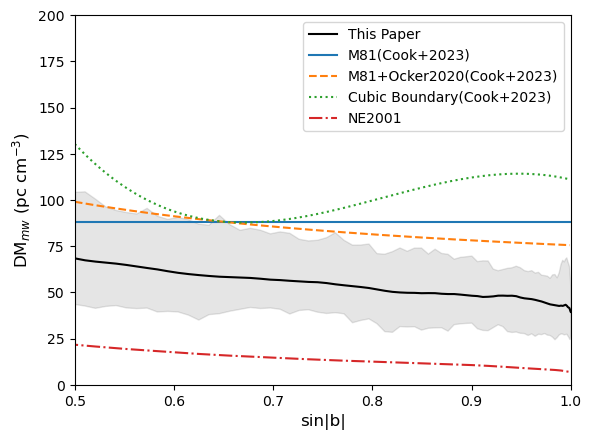}
    \caption{The relationship between \dmmw{} and latitude in our model is compared to the three \dmmw{} upper limit models from \citet{Cook2023}. The solid black line represents the mean \dmmw{} as a function of $\sin|b|$ in our model, while the shaded region indicates the scatter at the corresponding latitude. Note that the DM contributions from the LMC and SMC halos are excluded from this analysis, as CHIME FRBs cannot constrain \dmmw{} in these regions, which exhibit higher DM compared to the rest of the sky. The solid blue line, dashed orange line, and dotted green line correspond to the three upper bound models of \dmmw{} (models (i), (ii), and (iv) respectively) from \citet{Cook2023}. The dashed red line represents \dmmwism{} from NE2001 as a function of latitude.}
    \label{fig:cook_comp}
\end{figure}

Meanwhile, \citet{Cook2023} presents four upper bound models for the DM contribution from the MW (\dmmw{}) (including both the disk and halo components) for $|b| \in (30^\circ, 90^\circ)$, utilizing the lower bounds from CHIME FRBs' DM. The models are: (i) A uniform \dmmw{} across the sky, determined by the DM of FRB 20200120E, an FRB associated with M81, with $\dm=87.8\dmunits$ \citep{Bhardwaj2021}; (ii)A constant \dmmwhalo{} across the sky based on FRB 20200120E, combined with the MW disk model from \citet{Ocker2020}; (iii) A model fitting \dmmw{} using locally weighted scatterplot smoothing (LOWESS; \citet{Cleveland1979RobustLW}); and (iv) A model treating \dmmw{} as a polynomial boundary regression function of latitude. Since model (iii) provides the loosest constraint on the upper limit of the \dmmw{} model, we compare our model only with the other three constraint models and present the results in Fig.\ref{fig:cook_comp}. The figure shows that our model is consistent with the M81+Ocker2020 and the Cubic Boundary estimated by \citet{Cook2023}.

As another application of our Local Universe DM model, we estimate the host DM contribution for five literature FRBs with $z<0.028$ in Table \ref{tab:lowz_dm_host}, including the aforementioned FRB 20220319D \citep{Ravi2023}. The total host DM is calculated by subtracting contributions from the MW, the IGrM, the IGM, and foreground intervening halos. The host galaxy halo contribution, DM$_{\rm{host,halo}}$, is estimated as half of the \dmhalo{} contributed by the host galaxy. 
This adjustment is made because the FRBs are located within their host halos, meaning their sightlines only pass through part of the halo rather than traversing the entire extent. Here we assume that the FRB is located at the center of its host halo.
In other words, \texttt{pyhesdm} does not `know' about host halos and treats them as foreground halos.
The remaining unknown contribution, DM$_{\rm{host,unk}}$, is determined by subtracting DM$_{\rm{host,halo}}$ from the total host DM. 
We find that FRBs 20200120E \citep{Bhardwaj2021} and 20220319D are consistent with having a zero $\mathrm{DM_{host,unk}}$ within our uncertainties. 
FRB 20200120E, notably, is the nearby FRB found to lie within the outer disk of M81. 
The low DM$_{\rm{host,unk}}$ suggests that the FRB is not associated with any ionized nebulae within this galaxy.

\citet{Shin2024} recently examined the sightline of FRB20200723B and placed constraints on the gas properties of the intervening ``W-M sheet'' structure \citep{kim+2016_virgocluster}. This FRB was localized to NGC 4602 at a distance of $\sim33$ Mpc and has a total measured dispersion of $\dmfrb=244~\dmunits$. Using \texttt{pyhesdm}, we estimate $46~\dmunits$ are from the Milky Way ISM and halo. From the NEDLVS, we find that the host halo, NGC 4602, is the only significant halo contributor, with an estimated DM of $11~\dmunits$ (assuming an extent of 1 virial radii). No intersecting foreground clusters or groups are found within the Tully Catalog, as already mentioned by \citet{Shin2024}. \dmigm\ is estimated to be $6.6~\dmunits$ from our model. If one attributed all of \dmigm\ to the W-M sheet, the electron column density would be $\sim10^{19}\rm cm^{-2}$. Subsequently, this would imply an average electron number density of $\sim10^{-6} \rm~cm^{-3}$. Assuming complete ionization and that the filamentary gas contains 25\% Helium and 75\% Hydrogen by mass, this would imply a gas density of $\sim10^{-30}\rm~g~cm^{-3}$, which is roughly a few times the average cosmic baryon density at $z=0$ ($4\times10^{-31} \rm ~g~cm^{-3}$). Our estimate is consistent with the limits set by \citet{Shin2024} and additionally consistent with the expectation that cosmic web filaments are generally less than 10 times the average matter density of the universe \citep[e.g.][]{Cautun2013}. Our model would imply $\mathrm{DM_{host,unk}}/(1+z_{\rm frb})$ of $178~\dmunits$ (observed frame).
Together with FRB 20240210A (\citealt{Shannon2024}; see Table~\ref{tab:lowz_dm_host}) which has $\mathrm{DM_{host,unk}}/(1+z_{\rm frb}) =192\,\dmunits$, these two objects probably belong in the same category as the excess-\dmhost{} FRBs identified in \citet{Simha2023} and \citet{Lee2023}.

\begin{table*}
	\centering
	\caption{Host DM contribution for four FRBs with $z<0.028$. This table lists the Right Ascension (RA), Declination (Dec), redshift, and observed total DM (DM$_{\rm{tot}}$) for each FRB. The last two columns provide the estimated DM contributions from the host galaxy halo (DM$_{\rm{host,halo}}$) and the remaining unknown component (DM$_{\rm{host,unk}}$) in observed frame, respectively. DM$_{\rm{host,halo}}$ is taken as half the naive \dmhalo{} reported by \texttt{pyhesdm}, i.e.\ we assume that the FRB is at the center of the host halo. For FRB 20220319D, we adopt the \dmmwism{} value using the nearby PSR J0231+7026 instead of NE2001.}
	\label{tab:lowz_dm_host}
	\begin{tabular}{cccccccc} 
        \hline
        FRB & RA & Dec & $z_\mathrm{frb}$ & DM$_{\rm{tot}}$ & DM$_{\rm{host,halo}}/(1+z_\mathrm{frb}$) & DM$_{\rm{host,unk}}/(1+z_\mathrm{frb}$) & Reference \\
            & (deg) & (deg) &    & (\dmunits) & (\dmunits) & (\dmunits) & \\
        \hline
        20200120E & 149.48 & 68.82 & 0.0008 & 87.82 & 11 & 7 & \citet{Bhardwaj2021} \\
        20240210A & 8.78 & -28.27 & 0.0239 & 283.73 & 8 & 192 & \citet{Shannon2024} \\
        20220319D & 32.18 & 71.04 & 0.0112 & 110.95 & 15 &2 & \citet{Ravi2023} \\
        20181220A & 348.72 & 48.34 & 0.0275 & 209.40 & 16 & 63 & \citet{Amiri2024} \\
        20200723B & 190.16 & -5.14 & 0.0085 & 244 & 11 & 178 & \citet{Shin2024} \\
        \hline
	\end{tabular}\\
\end{table*}

We caution that the \hestia{}-based model for the MW halo and Local Group DM was run with a specific set of prescriptions for the subgrid physics --- namely the Auriga model --- that govern the feedback mechanisms impacting the CGM and IGrM of the Local Group.
This means that our model, especially for the MW halo and satellites, should be regarded as a prediction. 
Moreover, we emphasise that \hestia{} is not constrained to reproduce the satellite distribution in the Local Group. It is therefore possible that the ionised gas distribution in the inner MW halo as well as the LMC/SMC would be affected by the detailed MW satellite merger history which is not specifically modeled in \hestia{}.
This affects, for example, the gas distribution around the Magellanic Clouds, but there might also be unforeseen contributions by gas stripped during the merger of the hypothetical Sagittarius dwarf galaxy \citep{Tepper-Garcia2018}.

\section{Conclusions}\label{sec:conclusion}
In this paper, we construct a realistic DM model for the Local Universe beyond the Galactic disk, that is composed of three distinct layers, separated by proximity. The first layer is the DM model for the MW halo, in which we compute the DM contribution from the MW halo utilizing constrained hydrodynamic simulations from the \hestia{} simulation suite.
The \hestia{} simulation is designed to produce halo pairs representing the MW and M31 with relative distance matching the observation, with both galaxies exhibiting disk and bulge scales that match observational data. 
As a hydrodynamical simulation with a full supernova and AGN feedback prescription, \hestia{} predicts not only the gas distribution within the MW halo but also in the IGrM within the Local Group. 
We modeled the position of the Sun by ensuring it resides within the MW disk, at the correct distance from the Galactic Center, and that M31 is positioned at the correct Galactic longitude. 

The second layer is the DM model of the Local Group ($D<3.4\Mpc$), which we also derive from \hestia{}. In this component, the IGrM shell is rotated to ensure that M31 is positioned at the correct Galactic latitude. To improve accuracy, we remove the contribution from satellite galaxies that appear in the simulation but do not exist in the real universe. Additionally, we incorporate three dwarf galaxies—LMC, SMC, and M33—into the IGrM model by identifying simulated counterparts within the three highest-resolution hydrodynamic simulations in \hestia{}. We select those with stellar masses within 25\% of the observed values and with distances to their host galaxies that match the observations.

The outer layers represents the DM model for the Local Universe beyond the Local Group (3.4 Mpc $<D<120$ Mpc). We compute the DM contributions from the IGM and halos separately, utilizing the detailed spectroscopic observations that have been compiled in the Local Universe over the past decades. The DM model for the IGM is derived using the \hamlet{} density field computed from the Cosmicflows-4 catalog of peculiar velocities, while the DM model for galaxies and clusters is calculated using data from the NEDLVS catalog and the Tully15 group catalog. For each sightline, we identify relevant galaxies and clusters, modeling the gas profiles within the halos using the mNFW profile.
The \dmigm{} and \dmhalos{} are parametrised by free parameters governing the relative baryon fractions in the IGM as well as CGM/IGrM/ICM gas fractions across different halo masses.
We therefore expect our Local Universe \dmigm{} and \dmhalos{} model to have sufficient flexibility for general use, although in the future FRB samples might get large enough to reveal discrepancies in the modified NFW profiles adopted here.

We briefly compare our MW DM model with constraints on the Milky Way halo from previous studies by \citet{Ravi2023} and \citet{Cook2023}, demonstrating that our model is consistent with these previous studies.
As a first application, we apply our model to 5 localized FRBs that fall within $D<120\,$Mpc to study their host contributions.
We identify one new `naked' FRB (FRB 20200120E) with host contributions consistent with zero in addition to the previously-studied FRB 20220319D, while two objects (FRB 20240210A and FRB 20200723B) have $\dmhost \sim 200\,\dmunits$ that would lie on the tail of the \dmhost{} distribution.

Our DM model is available as a publicly-available Python package: \texttt{pyhesdm}, which is released on GitHub. This Python package provides functions that calculate the both the combined DM contribution derived in this paper, as well as the individual components (the Milky Way halo, the Intra-group Medium, the IGM, or halos) for a given Galactic latitude and longitude. These functions also allow users to adjust free parameters, including \figm{} (the fraction of cosmic baryons in the diffuse IGM) and the parameters in the mNFW model (\fgas, $y_0$ and $\alpha$). 

The main purpose of this model is to improve the constraints of the cosmic baryon distribution analyzed from the DM of localized FRBs \citep[see][]{Khrykin2024a}. 
We hope it will support all similar, future FRB 
analyses, especially low-$z$ FRB surveys like BURSTT \citep{Lin2022}, 
where variations in the Local Universe may
have significant impact.
This model can also optimize studies on constraining on cosmological parameters and galactic feedback utilizing \dmcosmic{} of localized FRBs.

\section*{Acknowledgements}

Kavli IPMU is supported by the World Premier International Research Center Initiative
(WPI), MEXT, Japan. We also acknowledge the use of \texttt{iDark} Cluster provided by Kavli IPMU for offering abundant computing resources, which was essential to our model calculations. YH gratefully acknowledges support from the Global Science Graduate Course (GSGC) program at the University of Tokyo. SS is grateful for the short-term fellowship from the Japan Society for the Promotion of Science (JSPS; award PE24009) to work at Kavli IPMU during the analysis phase of this project. SS is currently a Brinson postdoctoral fellow jointly at Northwestern University and the University of Chicago.
\section*{Data Availability}

This study uses data from the NEDLVS catalog (DOI: 10.26132/NED8). The \texttt{pygedm} algorithm is utilized in this study (DOI: 10.1017/pasa.2021.33)

The DM model introduced by this study is available as a publicly-available Python package on GitHub: \href{ https://github.com/yuxinhuang1229/pyhesdm}{ https://github.com/yuxinhuang1229/pyhesdm}.



\bibliographystyle{mnras}
\bibliography{local_dm} 




\appendix

\section{Validation of DM Calculation}

We validate our DM calculation code by computing the IGM contribution to the DM (\dmigm) at $z=0.1$ using the TNG100-1 simulation. The sightline that origin at $z=0.1$ can be divided into nine segments, corresponding to snapshot 91 through snapshot 99. The redshift of each snapshot and the length of the segment within each snapshot are listed in Table \ref{tab:snap_z}. The segments are divided such that the redshift at the starting point of the segment in snapshot $i$ $z_{i,\rm{start}}=(z_i+z_{i+1})/2$ where $z_i$ is the redshift corresponding to snapshot $i$ ($i=91$,...,98) with $z_{99,\rm{start}}$ = 0. Then the length of each segment within snapshot $i$ ($l_i$) is then calculated by the comoving distance from $z_{i,\rm{start}}$ to $z_{i-1,\rm{start}}$ ($i=91$,...,99).

\begin{table}
	\centering
	\caption{This table list the corresponding redshift of snapshot 91 through 99, along with the redshift at the starting point of the line segment inside the snapshot, and the length of the segment.}
	\label{tab:snap_z}
	\begin{tabular}{cccc} 
		\hline
		snapshot  & $z$ & ray start $z$ & ray length \\
		\hline
		99 & 0 & 0 & 21.046 Mpc \\
		98 & 0.009522 & 0.004761 & 52.786 Mpc \\
        97 & 0.023974 & 0.016748 & 52.986 Mpc \\
        96 & 0.033724 & 0.028849 & 53.433 Mpc \\
        95 & 0.048524 & 0.041124 & 53.621 Mpc \\
        94 & 0.058507 & 0.053515 & 54.058 Mpc \\
        93 & 0.073661 & 0.066084 & 54.233 Mpc \\
        92 & 0.083884 & 0.078773 & 54.659 Mpc \\
        91 & 0.099402 & 0.091643 & 35.302 Mpc \\
        90 & 0.109870 & 0.1 & / \\
		\hline
	\end{tabular}
\end{table}

\begin{figure}
    \centering
	\includegraphics[width=\columnwidth]{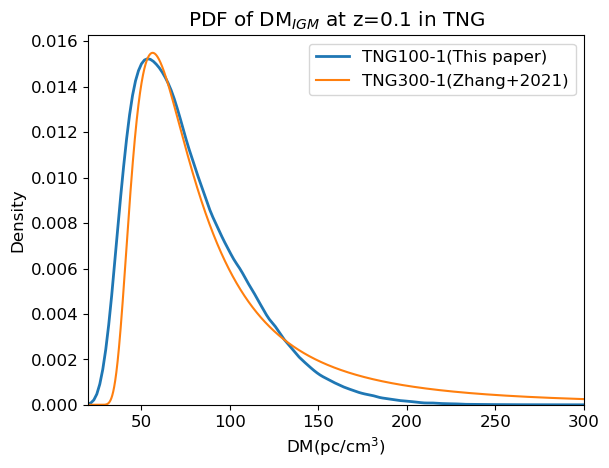}
    \caption{The distribution of \dmigm{} at redshift 0.1, calculated using TNG100-1 with the script from this paper (blue), comparing with the model of this distribution derived from TNG300-1 in \citet{Zhang2021} (orange).}
    \label{fig:dmigm_dist}
\end{figure}

We randomly select 100 sightlines in each snapshot with lengths as shown in Table \ref{tab:snap_z}. We calculate their DM with the method shown in Sec.\ref{subsec:mwhalo}, and then randomly combine the sightlines from snapshots 91 through 99 to generate $10^6$ different combinations, representing $10^6$ samples of \dmigm{} at $z=0.1$.The DM distribution of these combinations is shown in Fig.\ref{fig:dmigm_dist}, and is compared with the \dmigm{} distribution model $z=0.1$ derived from TNG300-1 in \citet{Zhang2021}. Due to differences in the simulations and the methods used, there is a slight discrepancy between the two results. However, they generally agree with each other, validating the accuracy of our method.

\section{Other Simulated Milky Ways in \hestia{} Simulation}

\begin{figure}
    \centering
	\includegraphics[width=\columnwidth]{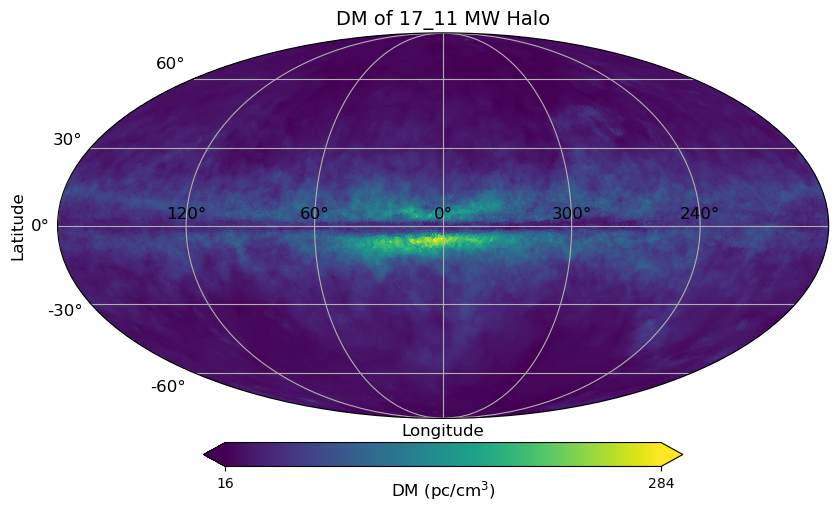}
        \includegraphics[width=\columnwidth]{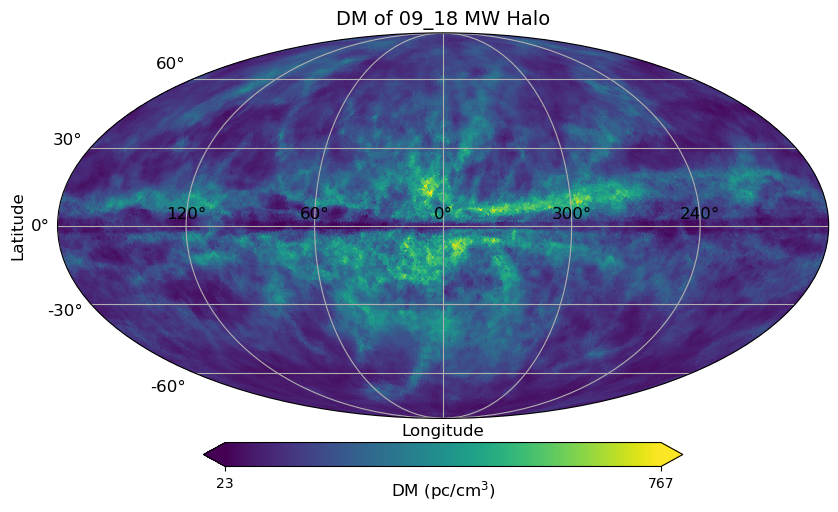}
    \caption{The \dmmwhalo{} model (without the disk component) computed from the simulated MWs in the 17\_11 box (upper) and the 09\_18 box (lower). The disk model is the same as we used in Section \ref{subsec:mwhalo}.}
    \label{fig:dm_mwhalo_compare}
\end{figure}

\begin{figure}
    \centering
	\includegraphics[width=\columnwidth]{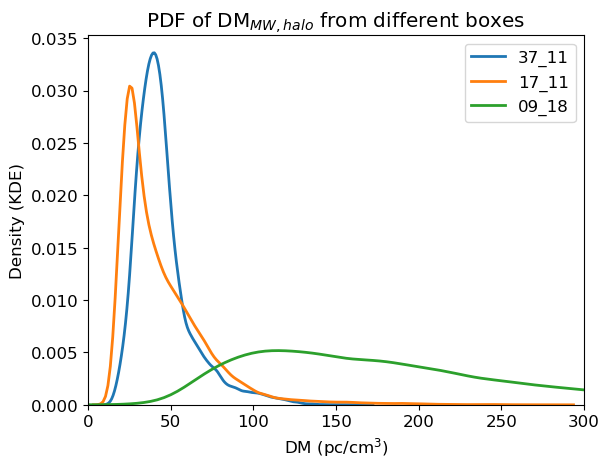}
    \caption{PDF of \dmmwhalo{} for the simulated Milky Ways in the three highest-resolution boxes: 37\_11 (blue), 17\_11 (orange) and 09\_18 (green).}
    \label{fig:dm_mwhalo_pdf_compare}
\end{figure}

In this section, we present the \dmmwhalo{} models derived from the simulated Milky Ways in the two other highest-resolution boxes of the \hestia{} simulation: 17\_11 and 09\_18. These two simulated MWs were not selected for our DM model because their bulge effective radii and disk scale lengths are more than twice the observed values reported in \citet{Bland-Hawthorn2016}. However, we include the \dmmwhalo{} of these two models in Fig.\ref{fig:dm_mwhalo_compare} for comparison with the model adopted in our DM calculations. The disk model used for calculating these \dmmwhalo{} values is the same as that in Section \ref{subsec:mwhalo}, defined by a cylinder with $r=20$\,kpc and $h=1$\,kpc. The PDFs of \dmmwhalo{} for all three simulated MWs are shown in Fig.\ref{fig:dm_mwhalo_pdf_compare}. The \dmmwhalo{} calculated from the 17\_11 box has a mean value of $\langle\dmmwhalo\rangle=44\,\dmunits$, a standard deviation of $\sigma_{\rm{MW,halo}}=27\,\dmunits$, a median value of $36\,\dmunits$ and a NMAD of $12\,\dmunits$, which is similar to the \dmmwhalo{} calculated from the 37\_11 box. However, the Milky Way in the 09\_18 box stands out as an outlier among the three simulated MWs. It exhibits a DM range from 23\,\dmunits{} to 767\,\dmunits, with a relatively flat PDF over this range. By adjusting the radius and height of the disk model, we ruled out the possibility that the large DM contribution originates from the ISM, and instead found that the MW halo contains diffuse gas with relatively high density, which explains its unusual \dmmwhalo{} map and PDF.

\bsp	
\label{lastpage}
\end{document}